\documentclass[prd,twocolumn,showkeys]{revtex4}
\usepackage{amssymb}
\usepackage{amsmath}
\usepackage{graphicx}
\usepackage{slashed} 
\usepackage{psfrag}

\newcommand{\fieldtensor}{\ensuremath{\mathcal{G}}}
\newcommand{\upperbub}{\ensuremath{\mathcal{U}}}
\newcommand{\hardbub}{\ensuremath{\mathcal{H}}}
\newcommand{\gluebub}{\ensuremath{\mathcal{L}}}
\newcommand{\gluedist}{\ensuremath{f}}
\newcommand{\qqbar}{\ensuremath{q \bar{q}}}

\newcommand{\pdfamp}{\ensuremath{\mathcal{L}}}
\newcommand{\picineq}[2]{ \ensuremath{\begin{array}{c} \includegraphics[scale=#2]{#1} \end{array} } }
\newcommand{\trans}{\ensuremath{t}}
\newcommand{\njet}{\ensuremath{n_{\rm J}}}

\newcommand\3[1]{{\bf #1}}

\begin{document}
\title{The Gluon Distribution Function and Factorization in Feynman Gauge}
\author{J.C. Collins} 
\author{T.C. Rogers}
\affiliation{Department of
Physics, Pennsylvania State University,\\ University Park, PA  16802,
USA}
\date{September 2, 2008}
\begin{abstract}
  A complication in proving factorization theorems in Feynman gauge is
  that individual graphs give a super-leading power of the hard scale
  when all the gluons inducing the hard scattering are longitudinally
  polarized. With the aid of an example in gluon-mediated 
  deep-inelastic scattering, we show that, although the super-leading terms
  cancel after a sum over graphs, there is a residual non-zero leading
  term from longitudinally polarized gluons.  This is due to the
  non-zero transverse momenta of the gluons in the target.  The
  non-cancellation, due to the non-Abelian property of the gauge
  group, is necessary to obtain the correct form of the gluon
  distribution function as a gauge-invariant matrix element.
\end{abstract}
\keywords{QCD, factorization}

\maketitle

\section{Introduction}
Many studies of the factorization 
theorems of perturbative quantum chromodynamics (pQCD) were performed
in axial or light-cone gauge. 
These gauges are appealing because there are no unphysical gluon polarizations or Faddeev-Popov ghosts, and the 
number density interpretation of parton distribution functions (PDFs) is clearer in light-cone gauge than 
in covariant gauges.
However, in axial or light-cone gauge, the gauge-field propagator has
unphysical singularities, which obstruct contour deformations needed
in the proofs of factorization.
Thus it is necessary to examine factorization 
in Feynman gauge where analytic properties associated with relativistic causality are manifest.  
A classic example of this is in 
the
factorization proof for the Drell-Yan 
process~\cite{CSS.proof,Bodwin}.
More recently, the process-dependent Wilson line directions in spin-dependent processes have been 
shown to be important for obtaining the correct relative sign in
single-spin asymmetries~\cite{Collins:2002kn,Brodsky:2002cx}; the
derivation of the Wilson line directions is easiest in Feynman gauge. 

An essential complication in Feynman gauge is that exchanges of
longitudinally polarized gluons between a collinear subgraph and the
hard-scattering subgraph contribute at one higher power of the hard
scale than in axial gauge.  One consequence is that there is no
suppression for adding extra gluon exchanges between the target 
and hard-scattering subgraphs.  A second and severe
consequence is that when a hard scattering is induced by gluons, we
find ``super-leading'' contributions, graph-by-graph.  For example,
super-leading terms for the deep inelastic structure function $F_2$
are of order $Q^2$, compared with the normal scaling behavior of $Q^0$
(up to logarithms of $Q$).  Arguments using gauge invariance are
needed to show that the super-leading terms cancel after a sum over
graphs for the hard scattering, and that the leading terms combine to
form gauge-invariant parton densities times conventional 
hard-scattering coefficients.  Unfortunately the details are not clearly
worked out in the literature.  That there is a cancellation of
super-leading terms was shown by Labastida and Sterman
\cite{Labastida:1984gy}.  They concentrated on issues in hadron-hadron
scattering, but their argument applies more generally, for example to
deep-inelastic lepton scattering (DIS).

However, their argument also appears to suggest that there is a
vanishing of the sum over graphs 
in which all exchanged gluons are longitudinally polarized.  In fact, as we will show in this
paper, the sum is actually nonzero; only the super-leading part
cancels.  This is a specific property of a non-Abelian gauge theory;
the sum is indeed zero in an Abelian theory.  
This is important because the non-zero sum
combines with terms with other polarizations for the gluons to give
the correct gauge-invariant form \cite{Collins:1981uw} for the gluon
density.

As far as we know, there is no detailed treatment in the literature
of the role played by the gluon polarization in factorization,
even at the level of low order Feynman graphs.  In this paper, we give
a detailed discussion at the level of two-gluon exchange (in the
amplitude).  Understanding the exact nature of the contributions from
the various kinds of polarization is important, not only to make sure
that ordinary factorization is understood, but also to allow correct
generalizations of factorization to be obtained (to include unintegrated PDFs, for instance). 

The key observation is that the target gluons are only approximately
collinear --- they have small but non-vanishing transverse components
of momentum relative to their parent hadron.  In contrast, the
Labastida-Sterman argument treated the gluons as having zero
transverse momentum.  This was sufficient to show cancellation of the
super-leading terms.  But it does not allow a correct treatment of the
remainder, which is power suppressed relative to the super-leading
terms, but does contribute to the leading power.

It should be emphasized that these issues are of direct practical importance.
Perturbative calculations often require the accurate identification of a gluon PDF.
In a subtractive formalism, perturbative corrections to the PDF are needed to obtain 
correct higher order hard-scattering coefficients.     
In different formalisms it is important to ensure that the same gluon PDF is 
being used.
In the pQCD dipole picture, for example,  the $q \bar{q}$ dipole cross section depends on the 
gluon PDF in the target~(see \cite{Ewerz:2007md} and references therein).  Ideally, it should be possible to show that this is the same 
gluon PDF that arises in standard pQCD factorization theorems. 

\begin{figure*}
\centering
  \begin{tabular}{c@{\hspace*{5mm}}c}
    \includegraphics[scale=0.4]{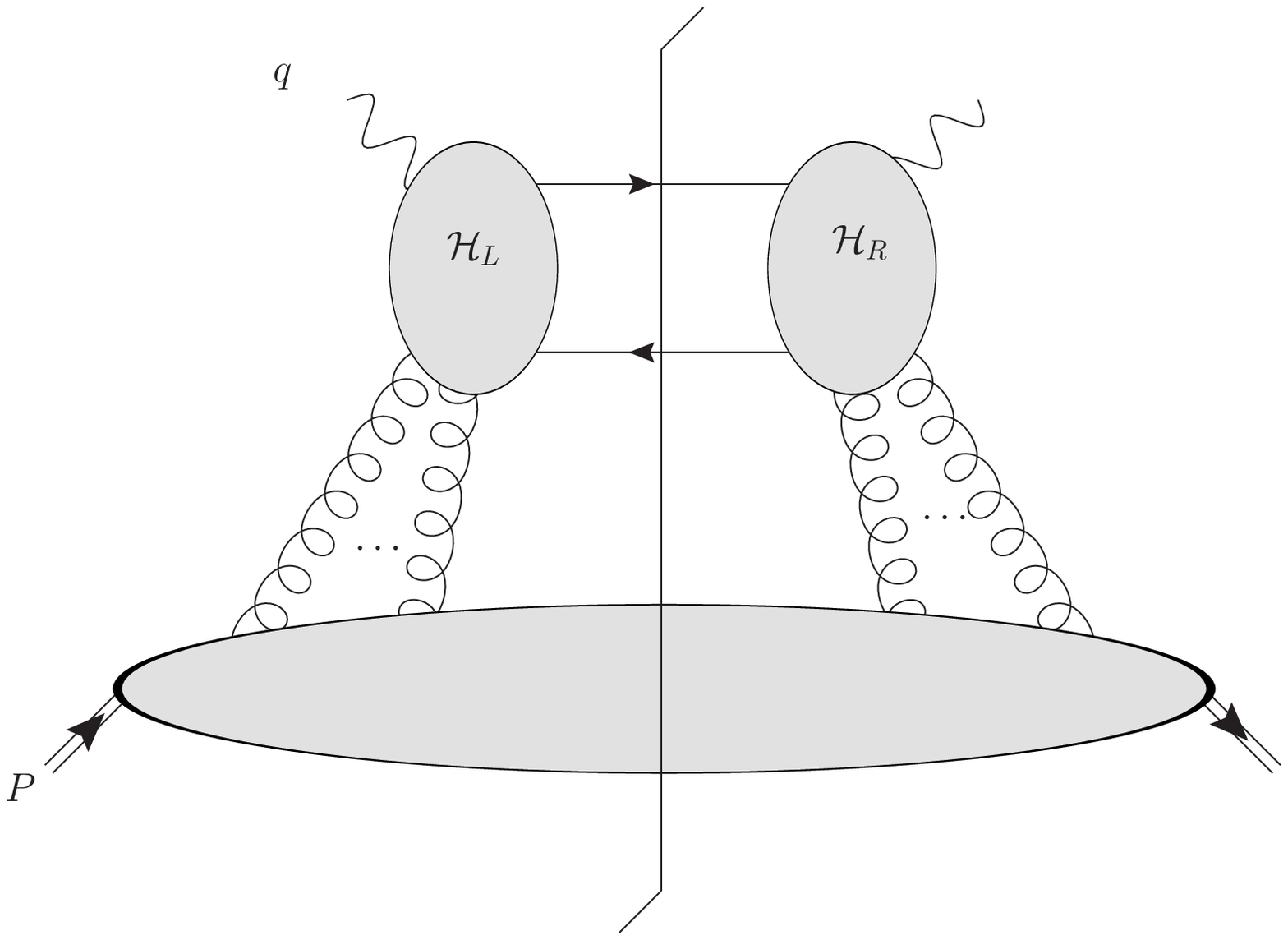}
    &
    \includegraphics[scale=0.4]{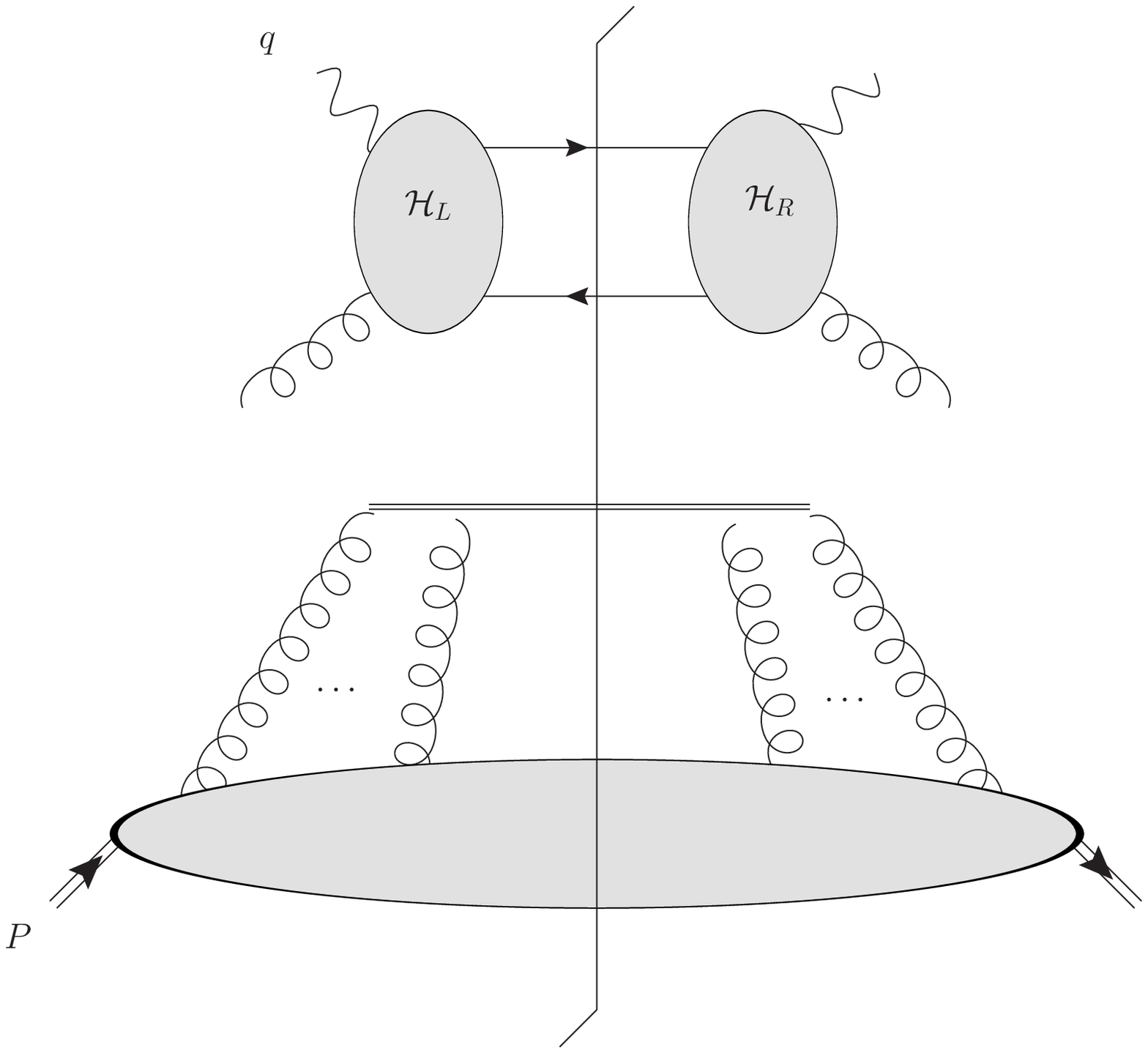}
    \\
    (a) & (b)
   \end{tabular}
\caption{(a) Leading region for $\gamma^{\ast} p \to \qqbar X$. (b) Graphical structure obtained after applying Ward identities to the target-collinear lines in (a).}
\label{fig:factorized}
\end{figure*}

The basic steps for deriving factorization for inclusive $\gamma^{\ast} p \to X $ scattering are as follows:
\begin{enumerate} 
\item 
Consider the most general type of region that
contributes at or above the standard leading power in $Q/\Lambda$.
Here, $\Lambda$ is a typical hadronic mass.
In this paper, we will only consider the contribution illustrated by the 
cut diagram in Fig.~\ref{fig:factorized}(a).  
The hard scattering is induced by gluons from 
the lower bubble, $\gluebub$, all of whose lines are collinear to the 
target.  
The final state of the hard scattering consists of a quark and
antiquark, which emerge from the hard-scattering bubbles, $\hardbub_{L/R}$ 
at wide angles.
Arbitrarily many target collinear gluons (represented by the gluons and the ``$\cdots$'')
may attach the target bubble to the hard bubbles.
In general, the outgoing quark and anti-quark evolve into bubbles of 
final-state collinear lines.  However, the integration contour in the sum over final-state momentum
may be deformed into the complex plane 
to where the final-state lines may be treated as being off-shell by
order $Q^2$ \cite{Labastida:1984gy}.
\item  
Use Ward identities and appropriate approximations 
to disentangle and factorize all the ``extra'' gluons.  
For the set of graphs we consider,
the result should be a convolution
product of the on-shell amplitude for $\gamma^{\ast} g \to \qqbar$ 
scattering with a non-perturbative factor containing target-collinear gluons. 
A graphical representation of the result of this last step is shown in Fig.~\ref{fig:factorized}(b). 
The double lines represent Wilson lines that make the parton densities
gauge invariant.
\item 
The non-perturbative factor in 
the final factorization formula of step 2 indicates an appropriate operator 
definition for the gluon PDF.  
This same operator definition appears in other 
factorization formulas for other processes, thus allowing the gluon density to be
fitted and used in multiple calculations. 
\end{enumerate}

We will follow these steps for the case of one and two gluons.
In a complete derivation of factorization, we also need 
to perform double counting subtractions (not shown explicitly in
Fig.~\ref{fig:factorized}).
However, the details of the subtraction procedure are not important
for the main issues discussed in this paper, which focuses only on the
factorization and identification of the gluon PDF for a specific
region of momentum space. 
We will therefore include only a brief discussion of subtractions, in
an appendix.

In Sect.~\ref{sec:setup}, we describe the kinematics of DIS and the notation and conventions to be
used in the rest of the paper.  
In Sect.~\ref{sec:one_gluon} we treat the 
simple case of a single target-collinear gluon, to establish our
basic technique.
In Sect.~\ref{sec:gluon_dist} we describe the properties of the standard integrated gluon distribution function. 
Then in Sect.~\ref{sec:2gluons} we extend the argument of Sect.~\ref{sec:one_gluon} to the case of two target-collinear gluons and 
illustrate the role the longitudinally polarized gluons play in the definition 
of the gluon PDF in Feynman gauge.  
We give comments and a
summary in Sect.~\ref{sec:conclusion}.
In an appendix we review the subtraction procedure appropriate for a
more general treatment.

\section{Basic Setup}
\label{sec:setup}
\begin{figure}
\centering
    \includegraphics[scale=0.4]{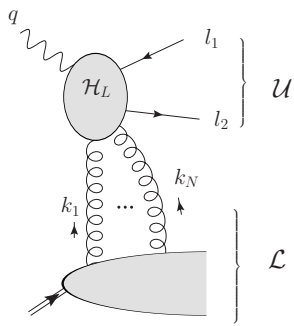}
\caption{Basic amplitude with arbitrarily many target collinear gluons.}
\label{fig:amplitude}
\end{figure}
The basic kinematic variables are the standard ones
for DIS.
The light-front coordinates of the momenta of the incoming proton and
photon are, respectively, 
\begin{equation}
P = \left( P^+,\frac{M_p^2}{2P^+}, {\bf 0}_t  \right), \qquad q = 
\left( -x P^+, \frac{Q^2}{2 x P^+}, {\bf 0}_t \right).
\end{equation}
Since we have kept the proton mass nonzero, the 
longitudinal momentum fraction, $x$, is not exactly 
equal to the usual Bjorken $x_{\rm Bj}$, but is related to it by
\begin{equation}
  x = \frac{2 \, x_{\rm Bj}}{1+\sqrt{ 1 + 4\, \frac{M_p^2}{Q^2}\, 
x_{\rm Bj}^2}}.
\end{equation}
We work in the Breit frame, where $x P^+ = Q/\sqrt{2}$.
To characterize the forward direction, we define the exactly light-like vector
\begin{equation}
\njet = (0,1,\3{0}_\trans).
\end{equation} 

In this paper we restrict our attention to contributions from graphs
of the form of Fig.~\ref{fig:factorized}(a), and 
it will be most convenient to work at the level of the amplitude shown in Fig.~\ref{fig:amplitude}.
The upper part of the graph, denoted by $\upperbub$, describes the 
scattering of a virtual photon off gluons.  
The lower part of the graph, denoted by $\pdfamp$, describes the 
emission of the gluons from the target proton.
In general, the amplitude for scattering off $N$ gluons in the target can be conveniently expressed as the contraction of  
$\upperbub$ and $\pdfamp$.
With the momentum labels shown in Fig.~\ref{fig:amplitude}, we have
\begin{equation}
\label{eq:amp}
M^\nu =  \upperbub(l_1,l_2;\left\{ k_j \right\} )^{\nu, \mu_1 \cdots \mu_N} 
  \; \pdfamp(P;\left\{k_j\right\})_{\mu_1 \cdots \mu_N}.
\end{equation}
Here $j$ runs from $1$ to $N$.  We will use $\nu$ to label the 
electromagnetic index, and $\mu_j$ to label the gluon indices.
The quark and anti-quark momenta are $l_1$ and $l_2$ and 
emerge from the hard scattering at wide angles.
The $k_j$ label the gluons connecting the two subgraphs.
We will be examining the contribution when each is in a neighborhood of the 
target collinear region:
\begin{equation}
\label{eq:ksize}
k_j \sim \left(Q, \frac{\Lambda^2}{Q}, \Lambda \right).
\end{equation}

We can treat both $\upperbub(l_1,l_2;\left\{ k_j \right\} )$ and
$\pdfamp(P;\left\{k_j\right\})$ as being obtained by a sum over all
graphs with the requisite external lines.  From the examination of
examples, it can be seen that each graphical contribution to the
amplitude $M^\nu$ is obtained $N!$ times. We compensate by defining
$\pdfamp(P;\left\{k_j\right\})$ to include a factor $1/N!$.

\section{Simplest Case: A Single Target Gluon}
\label{sec:one_gluon}

\begin{figure}
\centering
\begin{tabular}{c@{\hspace*{3mm}}c}
    \includegraphics[scale=0.4]{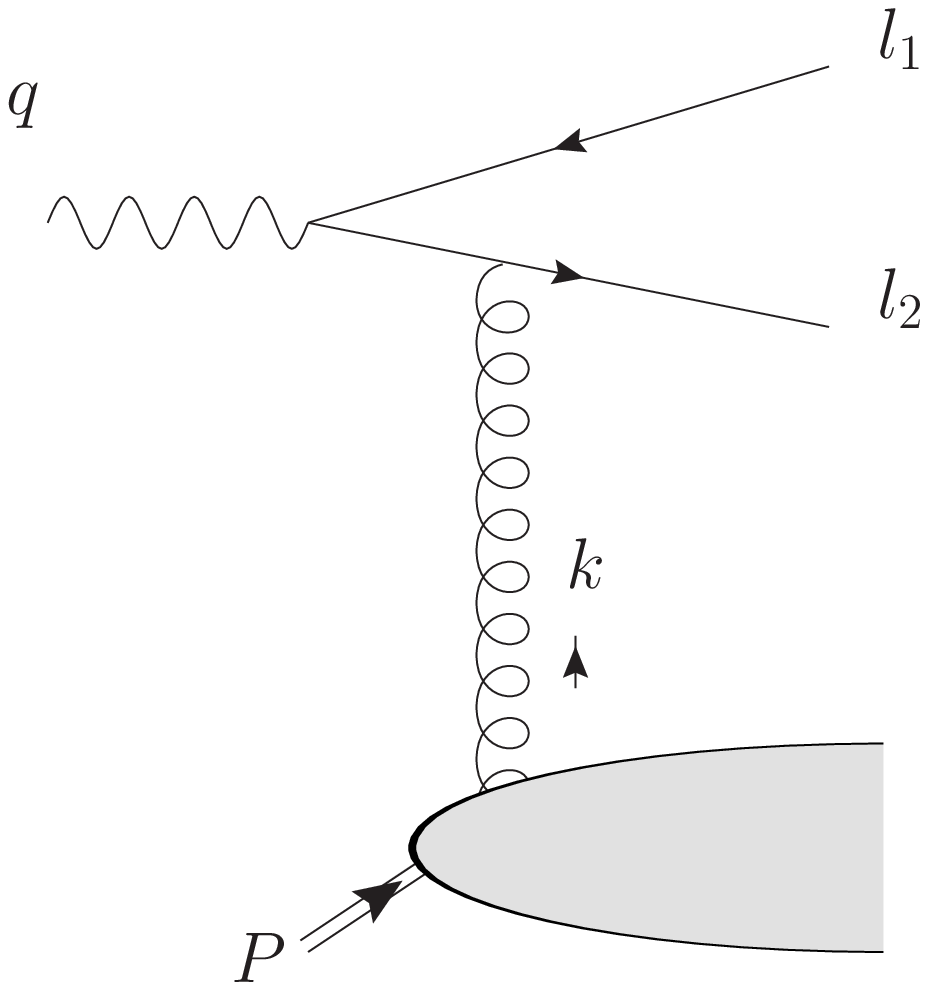} 
   & 
    \includegraphics[scale=0.4]{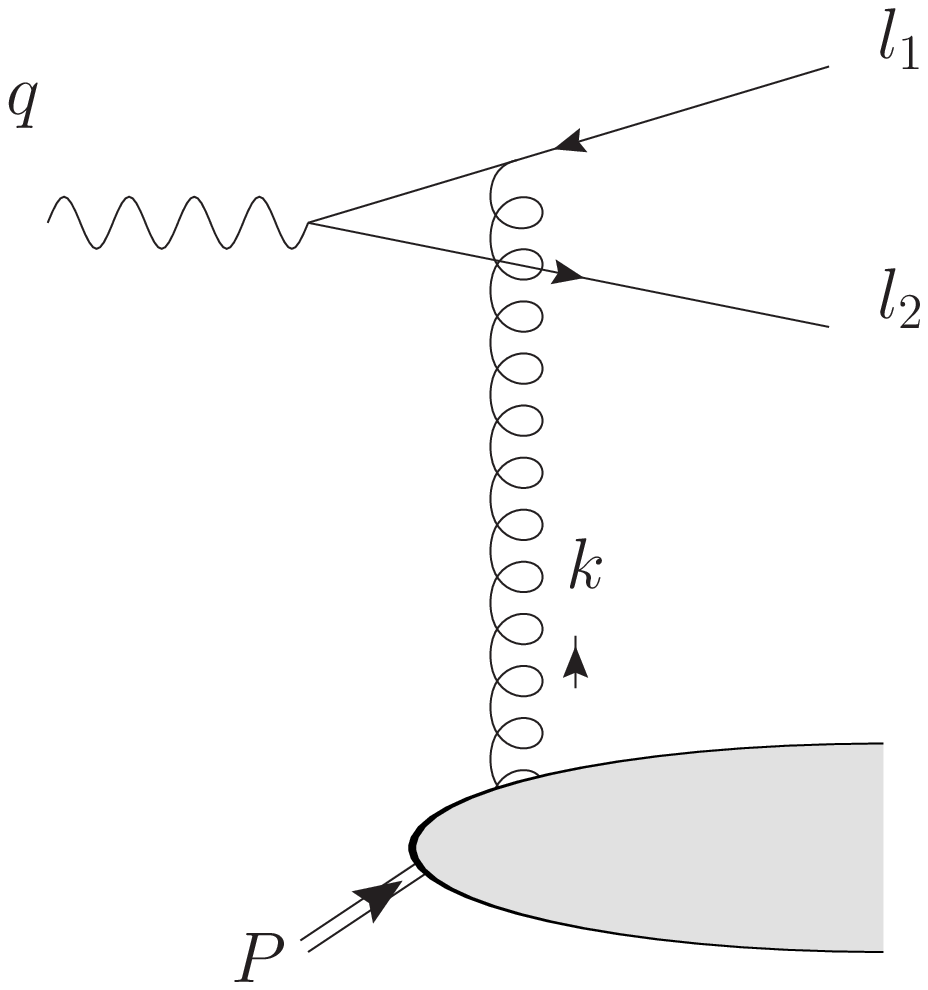}
   \\[3mm]
   (a) & (b)
\end{tabular}
\caption{Amplitudes for scattering off a single gluon.}
\label{fig:one_gluon}
\end{figure}
We start with the 
simple case of the exchange of one gluon, as 
shown in the graphs of Fig.~\ref{fig:one_gluon}. 
The amplitude is of the form
\begin{equation}
\label{eq:one_gluon}
M^{\nu}_{(1g)} = \upperbub^{\nu}_{\mu^{\prime}} g^{\mu^{\prime} \mu} \pdfamp_{\mu}.
\end{equation}
The subscript $(1g)$ denotes that one gluon is exchanged, and we have
explicitly shown the numerator $g^{\mu'\mu}$ of the gluon propagator.  
To simplify notation, we will often omit the momentum arguments and
the index $\nu$ of the electromagnetic current.

The gluon is collinear to the target, and $l_1$ and $l_2$ are at wide
angle, so that the quark subgraph is the hard scattering.  
We will show that, after a sum over diagrams, the lower part of the 
graphs in Fig.~\ref{fig:one_gluon}  can be separated into a factor that can be identified 
as a contribution  to the gluon distribution function.

There are regions where the target gluon may go far off-shell, but in that case it should be regarded 
as belonging to the hard-scattering subgraph at one higher power of
$g_s^2$.  For this paper, we restrict ourselves to 
the case where the hard-scattering coefficient is lowest order (LO) in $g_s^2$.
\subsection{Power counting}

To understand the power counting in $Q$, for large $Q/\Lambda$, we first
observe that the target is highly boosted in the plus direction, so that
\begin{align}
\label{eq:lowersize}
\pdfamp^+ \sim Q/\Lambda, \qquad \pdfamp_t \sim 1, \qquad \pdfamp^- \sim \Lambda/Q,
\end{align}
up to an overall power of $\Lambda$.  
Since $l_1$ and $l_2$ are at wide angles, all components 
of $\upperbub$ are of comparable size, and given by the dimension of
the quark part of the graph:
\begin{equation}
\label{eq:uppersize}
\upperbub^- \sim \upperbub^+ \sim \upperbub_\trans \sim Q^0.
\end{equation}
To obtain the structure function, we square the amplitude and
integrate over the final states.  The integral over the gluon momentum
$k$ is invariant under boosts from the target rest frame, and
therefore does not contribute any power of $Q$.  The Lorentz-invariant
phase space integral for the $q\bar{q}$ pair, 
\begin{equation}
  \frac{ d^3\3l_1 \, d^3\3l_2 }
       { (2\pi)^6 4 |\3l_1| |\3l_2| }
  (2\pi)^4 \delta^{(4)}(l_1+l_2-q+k),
\end{equation}
is dimensionless.  Thus it also contributes a zero power of $Q$ when
the 
quark-antiquark pair is at a wide angle, where all components of $l_1$
and $l_2$ are of order $Q$.  Thus the overall power of $Q$ is 
that for the squared amplitude.

From Eqs.~(\ref{eq:lowersize}), (\ref{eq:uppersize}), we see that the
largest power of $Q$ is from the term $\upperbub^-\pdfamp^+ \sim Q/\Lambda$,
for individual graphs.  Squaring this gives the previously mentioned
super-leading power of $Q^2$ in a structure function.  In the next subsection, we will use a
generalizable method to demonstrate cancellation of the super-leading
terms in the amplitude.  To give a correct treatment of the leading
terms ($Q^0$), we need to work 
at next-to-leading power accuracy in $Q$ relative to the super-leading power. We also need to show that
the leading terms correspond to a correct definition of the gluon
density.

\subsection{$K$ and $G$ gluons}

A convenient technique was introduced by 
Grammer and Yennie in the context of QED~\cite{Grammer:1973db}.
In this approach, the propagator numerator in Eq.~(\ref{eq:one_gluon}) 
is split into two parts:
\begin{equation}
\label{eq:gramyen1}
g^{\mu^{\prime} \mu} = K^{\mu^{\prime} \mu} + G^{\mu^{\prime} \mu},
\end{equation}
where
\begin{align}
\label{eq:gramyenK}
K^{\mu^{\prime} \mu} 
= &  \frac{k^{\mu^{\prime}} \njet^{\mu}}{k \cdot \njet},
\\
\label{eq:gramyenG}
G^{\mu^{\prime} \mu} = &  
g^{\mu^{\prime} \mu} 
- \frac{k^{\mu^{\prime}} \njet^{\mu}}{k \cdot \njet}.
\end{align}
In these definitions, 
we use the light-like direction
$\njet$ which 
characterizes
the outgoing jet direction.
The rationale for this choice is that the dominant term in
$\upperbub\cdot \pdfamp$ is the one 
obtained from the $g^{+-}$ component of the
gluon-propagator numerator. 
$K^{+-}$ reproduces exactly this component, the remaining components
of $K$ give smaller contributions, and $K$ has a factor $k^{\mu'}$,
which allows gauge invariance to be applied.

We use Eq.~(\ref{eq:gramyen1}) to separate Eq.~(\ref{eq:one_gluon})
into two parts which we call the ``$K$-term'' and the ``$G$-term.''
We will also use the terminology of the exchange of ``$K$-gluons''
and ``$G$-gluons''.

From Eq.~(\ref{eq:ksize}), we determine 
the $Q$ dependence that results from the individual components in the
$K$-term: 
\begin{align}
\upperbub^{-} K^{+-} \pdfamp^+ & \sim \frac{Q}{\Lambda}, \\
\upperbub^{t} K^{t-} \pdfamp^+ & \sim Q^0, \\
\mbox{All others} & \sim \frac{\Lambda}{Q}, \mbox{~or smaller}.
\end{align}
The $K^{+-}$ term gives the super-leading contribution, associated
with the power of $Q$ in $\pdfamp^+$.  Changing to $K^{t-}$ removes
one power of $Q$, by changing $k^+$ to $k^t$.  The resulting power,
$Q^0$, we identify as the standard leading behavior.

As we will see explicitly, gauge invariance ensures that 
$k$ dotted into the sum of all graphs in the upper bubble vanishes, 
so that
\begin{equation}
\label{eq:cancellation}
\sum_{\rm graphs} \upperbub_{\mu^{\prime}} K^{\mu^{\prime} \mu}  = 0.
\end{equation}
Hence the $K$-term can be dropped, so that Eq.~(\ref{eq:one_gluon})
gives
\begin{equation}
\label{eq:amp2}
M_{(1g)} = \upperbub_{\mu^{\prime}} G^{\mu^{\prime} \mu} \pdfamp_{\mu}.
\end{equation}
Only the terms with transverse components on $\mu'$ are unsuppressed in Eq.~(\ref{eq:amp2}).
This is easily verified by defining
\begin{equation} 
\label{eq:modpdfamp}
\tilde{\pdfamp}^{\mu^{\prime}} \equiv G^{\mu^{\prime} \mu} \pdfamp_{\mu}, 
\end{equation}
and recalling Eqs.~(\ref{eq:ksize}), (\ref{eq:lowersize}).
Checking each combination of indices we find,  
\begin{align}
\label{eq:tt}
\upperbub^t \tilde{\pdfamp}^t 
& \sim Q^0 \\
\upperbub^+\tilde{\pdfamp}^- & \sim \frac{\Lambda}{Q},
\label{eq:plusmin} \\
\upperbub^-\tilde{\pdfamp}^+ 
& = 0.
\label{eq:minplus}
\end{align}
Therefore, the transverse term dominates, and the other terms are
power suppressed or zero.

\subsection{Gauge-invariance calculation}
\label{sec:WI.basic}

We now give an explicit demonstration that $\sum_{\rm graphs}
\upperbub_{\mu'} K^{\mu'\mu} = 0$.
The graphs of Fig.~\ref{fig:one_gluon} give
\begin{widetext}
\begin{equation}
\label{eq:one.gluon}
\upperbub^{\mu^{\prime}}(k) = g_s t_\alpha \bar{u}(l_2) \left[ \gamma^{\mu^{\prime}} \left( \frac{1}{\slashed{l}_2 - \slashed{k} - m} \right) \gamma^{\nu}  + 
\gamma^{\nu} \left( \frac{1}{\slashed{k} - \slashed{l}_1 - m} \right) \gamma^{\mu^{\prime}} \right] v(l_1).
\end{equation}
Then for the $K$-term we have
\begin{equation}
\label{eq:subk}
\upperbub_{\mu^{\prime}} K^{\mu^{\prime} \mu} = g_s t_\alpha \frac{\njet^{\mu} }{k \cdot \njet} \bar{u}(l_2) \left[  \slashed{k} \left( \frac{1}{\slashed{l}_2 - \slashed{k} - m} \right) \gamma^{\nu}   \right. +  \left.
\gamma^{\nu} \left( \frac{1}{\slashed{k} - \slashed{l}_1 - m} \right) \slashed{k} \right] v(l_1),
\end{equation}
to which we apply the following identities for $\slashed{k}$:
\begin{align}
\slashed{k} & = - (\slashed{l}_2 - \slashed{k} - m) + (\slashed{l}_2 - m) && \text{in term 1}, \\
& =  (\slashed{k} - \slashed{l}_1 - m) + (\slashed{l}_1 + m)  && \text{in term 2}. 
\end{align}
Using the Dirac equation for
$\bar{u}(l_2)$ and $v(l_1)$, we find that two terms vanish and
the remaining terms exactly cancel, so
\begin{equation}
\upperbub_{\mu^{\prime}} K^{\mu^{\prime} \mu} = \frac{ g_s t_\alpha \njet^{\mu} }{k \cdot \njet} 
\left[-\bar{u}(l_2) \gamma^{\nu} v(l_1) + \bar{u}(l_2) \gamma^{\nu} v(l_1) \right] = 0.
\end{equation}
Therefore, only the $G$-term survives, so that
\begin{equation}
\label{eq:onegluon_fact}
M_{(1g)} =
\upperbub_{\mu^\prime} G^{\mu^\prime \mu} 
\pdfamp_{\mu} = 
\upperbub^{\mu} \tilde{\pdfamp}_{\mu}.
\end{equation}

We now have the power-counting 
of Eqs.~(\ref{eq:tt})--(\ref{eq:minplus}).  It is important to recognize that
we wrote the amplitude in the form of Eq.~(\ref{eq:onegluon_fact}) without making 
any approximations.
It is very tempting to replace $k$ by an exactly collinear value at
the start of the argument,
both in the upper, hard-scattering subgraph, and in Eqs.~(\ref{eq:gramyenK}) and
(\ref{eq:gramyenG}).  The approximated momentum is
on-shell and has zero transverse momentum.
This replacement is particularly natural to
make in the $K$ term, and was made by Labastida and Sterman
\cite{Labastida:1984gy}.  
However,
then we would have made an error
in each graph contributing to the $K$-term that is of order $k_t/Q$
\emph{relative} to the super-leading power $Q$.  That is, the error
would be of the same order 
as the leading terms in the final result.  
Using this kinematic
approximation in the $K$ term also entails using it in the $G$ term,
and we would not have obtained the contribution to $G^{\mu'\mu}$ involving
the transverse components of $k$. 
If we make the collinear approximation in the upper subgraph but not
in $K$, then the cancellation of the $K$-terms 
in Eq.~(\ref{eq:subk})
is no longer exact.
(The error would again be of order $k_t/Q$ relative to the super-leading power.)
These observations will be particularly important for obtaining the
correct two-gluon contribution in Sect.~\ref{sec:2gluons}. 

\subsection{Leading-power Factorization}
\label{sec:fact_steps}

The next step in
obtaining the factorization formula for scattering off a single target gluon is to drop power-suppressed terms. 
This means we can drop the contribution from Eq.~(\ref{eq:plusmin}) and keep only the sum over transverse components 
in Eq.~(\ref{eq:onegluon_fact}).  Furthermore, since we are
restricting to 
the region of $k$ given by Eq.~(\ref{eq:ksize}), we can 
now substitute the approximate parton momentum, 
\begin{equation}
\label{eq:k.hat}
\hat{k} = (k^+,0,{\bf 0}_t),
\end{equation}  
into the upper bubble \emph{only}. Thus
the hard scattering is initiated, as is usual, by a massless on-shell
parton of zero transverse momentum.  The amplitude is then
\begin{equation} 
\label{eq:psdropa}
M_{(1g)} = \sum_{j=1}^2 \upperbub^{j}(l_1,l_2;\hat{k}) \tilde{\pdfamp}_{j}(P;k)  + \mathcal{O}\left( \frac{\Lambda}{Q} \right),
\end{equation}
where the index $j$ runs only over the two transverse components.
In an unpolarized cross section or structure function, we square the amplitude and 
sum/integrate over final states. 
However, we have assumed above that the 
outgoing quarks are at wide angles, so that the collinear
approximation (\ref{eq:k.hat}) for the gluon can be safely applied in
the upper part of the graphs.  

But in the integral over \emph{all}
final states, there is also a 
leading
contribution from the region where one
of the intermediate quark  lines  is collinear to the target, i.e., 
$(k-l_1) \sim (k^+ - l_1^+,0,{\bf 0}_t)$ or $(k-l_2) \sim (k^+ - l_2^+,0,{\bf 0}_t)$.
This contribution to the cross section is already taken into account
at the parton-model level, and therefore an appropriate subtraction
should be made to avoid double counting,
according to the principles summarized in the Appendix.  
Thus the contribution of the graphs to the structure tensor takes the form
\begin{equation}
\label{eq:cross_section}
W^{\nu_2\nu_1} \sim \sum_{j,j^\prime} \int d\Pi \; 
   \upperbub^{\nu_1\,j}(l_1,l_2;\hat{k}) \upperbub^{\nu_2\, j^\prime \,\dagger}(l_1,l_2;\hat{k})\; 
   \tilde{\pdfamp}_j(P;k)
   \tilde{\pdfamp}^\dagger_{j^\prime}(P;k) - \text{subtraction terms} +
   \mathcal{O}\left( \frac{\Lambda}{Q} \right). 
\end{equation}
Here, $d \Pi$ denotes the complete integration measure including sums over final states.
The exact form of the subtraction terms is found by making 
the parton-model approximation on the struck quark, appropriate 
for the region of phase space where one of the quarks is
target collinear.
(The detailed steps for making this approximation are reviewed in the introductory sections of Ref.~\cite{Collins:2007ph}.)
We denote the parton-model approximation on the upper part of the graph 
by $T_{\rm PM} \upperbub^j(l_1,l_2;\hat{k}) \upperbub^{j^\prime \, \dagger}(l_1,l_2;\hat{k})$.
Furthermore, in the sum over final states for the unpolarized cross section, $\upperbub^j(l_1,l_2;\hat{k}) \upperbub^{j^\prime \, \dagger}(l_1,l_2;\hat{k})$ is diagonal in $j$ and $j^\prime$, 
so we may write Eq.~(\ref{eq:cross_section}) as,
\begin{equation}
\label{eq:cross_sectionb}
W \sim \int d\Pi \, \left[ \frac{1}{2}\sum_j (\hat{\upperbub}^j \hat{\upperbub}^{j \dagger} 
- T_{\rm PM}\hat{\upperbub}^j \hat{\upperbub}^{j \dagger})  \right] \left[ \sum_{j^\prime} \tilde{\pdfamp}_{j^\prime}(P;k) \tilde{\pdfamp}^\dagger_{j^\prime}(P;k) \right] + \mathcal{O}\left( \frac{\Lambda}{Q} \right),
\end{equation}
where $\hat{\upperbub}^j \equiv \upperbub^j(l_1,l_2;\hat{k})$.
We note that the manipulations needed to obtain Eq.~(\ref{eq:cross_sectionb}) 
can be made transparent by first writing the contractions of the upper and lower bubbles 
in terms of sums over transverse polarizations:
\begin{equation}
\sum_j \upperbub^j(l_1,l_2;\hat{k}) \tilde{\pdfamp}_j(P;k)  = \sum_{i,j} \sum_s \upperbub^j(l_1,l_2;\hat{k}) \left( \epsilon_{t,j} \right)^s \left( \epsilon_{t,i}  \right)^s  \tilde{\pdfamp}_i(P;k),
\end{equation}
where we have introduced transverse polarization vectors,
\begin{equation}
\left( \epsilon_{t,j} \right)^1 = (0,1,0,0), \qquad \left( \epsilon_{t,j} \right)^2 = (0,0,1,0).
\end{equation}
Then by the use of diagonality of 
$\upperbub^j(l_1,l_2;\hat{k})\upperbub^{j^\prime \,\dagger}(l_1,l_2;\hat{k})$
after the integral over the final state, the sums over polarizations 
produce Eq.~(\ref{eq:cross_sectionb}).

The first factor in the integrand of Eq.~(\ref{eq:cross_sectionb})
is just the unpolarized squared amplitude for $\gamma^{\ast} g \to q \bar{q}$ scattering minus a subtraction term.
For the rest of this paper, we will ignore the details of implementing 
the double counting subtraction since it only modifies the hard scattering but does not 
affect Ward identity arguments such as those of Sect.~\ref{sec:WI.basic}.  

Since $\upperbub^j(l_1,l_2;\hat{k})$ depends only on the plus component
of $k$ after the approximation of Eq.~(\ref{eq:cross_sectionb}), 
the integration over $k^-$ and ${\bf k}_t$ depends only on the target bubble.
Thus, we will ultimately identify the lowest order contribution 
to the gluon PDF with the factor,
\begin{equation}
\label{eq:oneg_pdf}
\sum_X k^+ \int \frac{d k^- d^2 {\bf k}_t }{(2 \pi)^4} \sum_{j^\prime} \tilde{\pdfamp}_{j^\prime}(P;k) \tilde{\pdfamp}^\dagger_{j^\prime}(P;k),
\end{equation}
where $\sum_X$ represents a general sum/integral over the final states of
the target bubble, and the factor $k^+$ gives the standard
normalization for the gluon density.
The factorization formula that follows from
Eq.~(\ref{eq:cross_sectionb}) can thus be expressed graphically as 
\begin{multline}
\label{eq:fact_oneg}
W \sim \int \frac{ d k^+ }{ 2k^+ } 
    \sum_j \left( \picineq{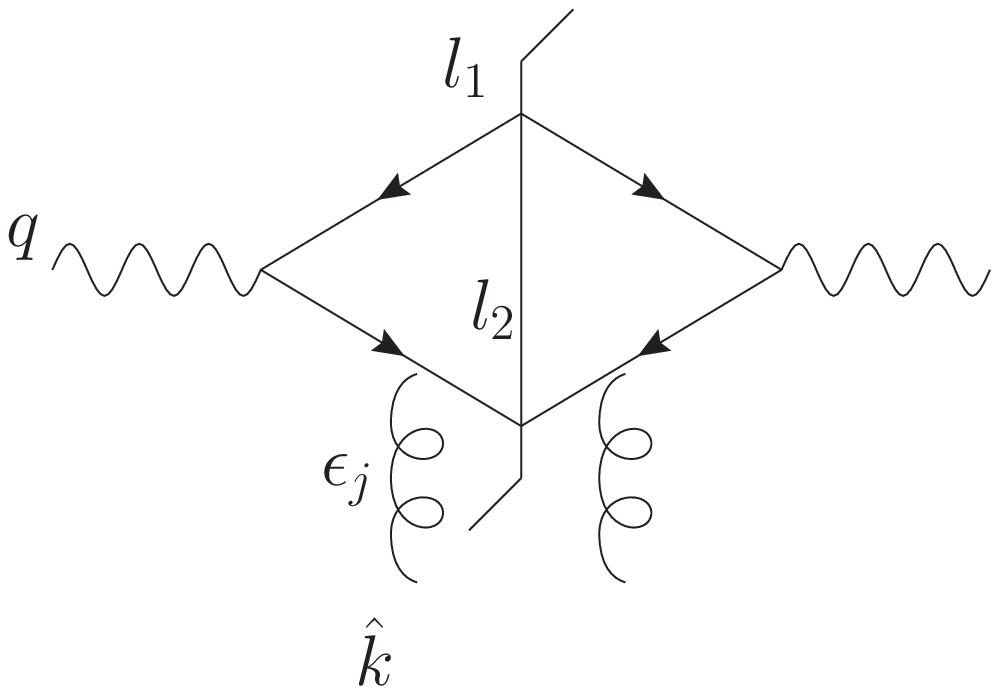}{0.28} + \picineq{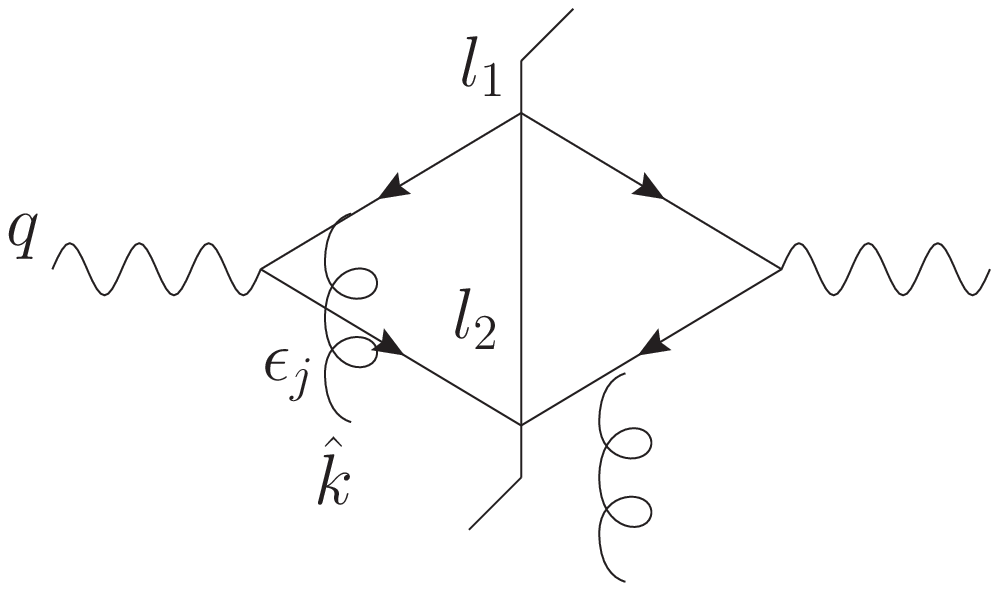}{0.28} + \picineq{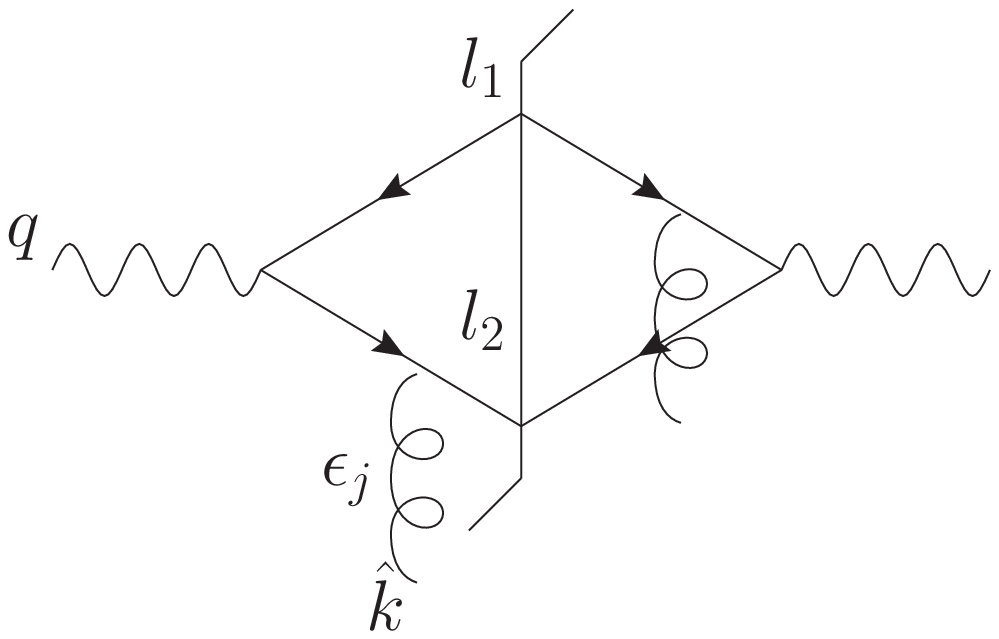}{0.28} + \picineq{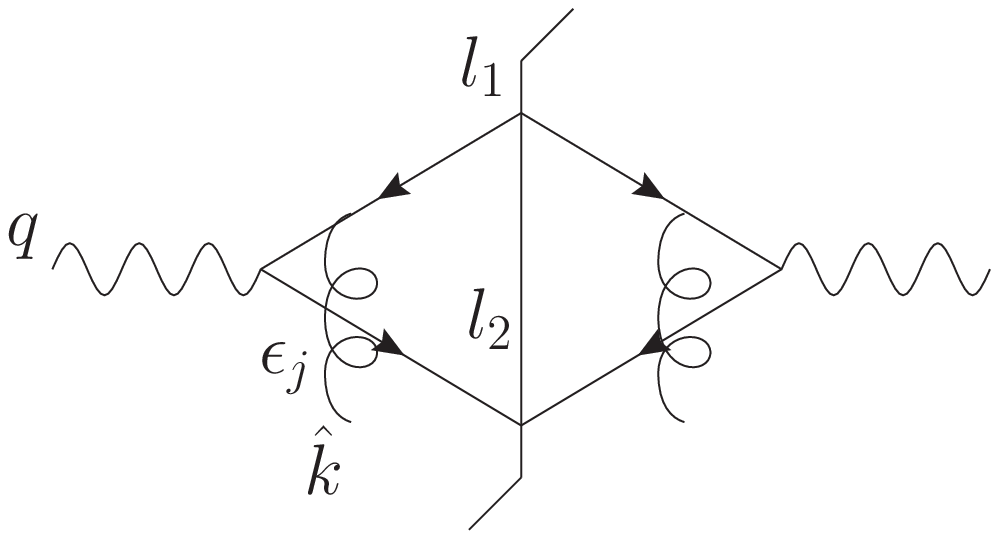}{0.28}  \right) \times \\ \times
\left(  \sum_X k^+ \int \frac{d k^- d^2 {\bf k}_t }{(2 \pi)^4} \sum_{j^\prime} \tilde{\pdfamp}_{j^\prime}(P;k) \tilde{\pdfamp}^\dagger_{j^\prime}(P;k)  \right) - \text{subtraction terms} + \mathcal{O}(\Lambda/Q).
\end{multline}
The upper part of the graph now has the interpretation of an on-shell scattering 
amplitude for a virtual photon to scatter off a transversely polarized gluon with 
exactly collinear momentum.
In a general treatment of factorization, the steps of this section 
need to be generalized to more than one gluon, and factors like Eq.~(\ref{eq:oneg_pdf}) need 
to be shown to correspond to an operator definition of the gluon distribution function.
Therefore, before we go further we must examine the general properties of the gluon distribution function.
\section{The Gluon Distribution Function}
\label{sec:gluon_dist}
\subsection{Operator Definition}
For Eq.~(\ref{eq:fact_oneg}) to be a consistent factorization 
formula for the single-gluon case, 
Eq.~(\ref{eq:oneg_pdf}) should 
correspond to the relevant contribution
to the operator definition of the gluon PDF. 
The standard definition~\cite{Collins:1981uw} of the integrated momentum space gluon distribution function for the proton is
\begin{equation}
\label{eq:gluedist}
\gluedist_{g/p}(\xi,P) \equiv \sum_{j=1}^2 \int \frac{dw^-}{2 \pi \xi P^+}    
e^{-i \xi P^+ w^- } \langle P | \; \fieldtensor^{+j} (0,w^-,{\bf 0}_t) \mathcal{P}_A  \fieldtensor^{+j}(0)  \; | P \rangle.
\end{equation}
Here $\xi= k\cdot\njet/P\cdot\njet = k^+/P^+$ is the longitudinal
momentum fraction of the proton carried by the target gluon.  
To make the definition exactly gauge invariant, we have inserted a
Wilson line operator in the adjoint representation, 
\begin{equation}
\mathcal{P}_A \equiv P \exp \left( -i g_s \int_0^{w^-} d y^- A^+_\beta(0,y^-,{\bf 0}_t) T_\beta \right),
\end{equation}
where the $T_\beta$ ($\beta = 1,2, \dots 8$) are the generators of SU(3) in
the adjoint representation and $P$ is the path-ordering operation.
The $T_\beta$ are related to the structure constants, $f_{\beta \gamma \kappa}$, by
$(T_\beta)_{\gamma \kappa} = -if_{\beta \gamma \kappa}$. 
The color indices of $\mathcal{P}_A$ are
contracted with those of the gauge field-strength tensor, whose
definition we recall:
\begin{equation}
\label{eq:fieldtensor}
\fieldtensor^{\mu \nu}_\alpha(z) = \partial^\mu A_\alpha^\nu(z) - \partial^\nu A_\alpha^\mu(z) - g_sf_{\alpha \beta \gamma} A_\beta^\mu(z) A_\gamma^\nu(z).
\end{equation}
Then examining how to apply Eq.~(\ref{eq:gluedist}) in perturbative
calculations yields Feynman rules 
for the gluon PDF.
In perturbative calculations, it is convenient to substitute the following 
identity for the Wilson line operator,
\begin{multline}
P \exp \left( -i g_s \int_0^{w^-} d y^- A^+_\beta (0,y^-,{\bf 0}_t) T_\beta \right) \\
= \left[ P \exp \left( -i g_s \int_0^{\infty} d y^- A^+_\beta (y^- + w^-) T_\beta \right) \right]^\dagger 
P \exp \left( -i g_s \int_0^{\infty} d y^{\prime -} A^+_{\beta} (y^{\prime -}) T_{\beta} \right),
\end{multline}
and to insert a complete sum over final states.  
Then Eq.~(\ref{eq:gluedist}) becomes 
\begin{multline}
\label{eq:gluedistb}
\gluedist_{g/p}(\xi,P) = \sum_X \sum_\alpha \sum_{j=1}^2 \int \frac{dw^-}{2 \pi \xi P^+}    
e^{-i \xi P^+ w^- } \langle P | \; \fieldtensor^{+ j}_\alpha (0,w^-,{\bf 0}_t) \left[ P \exp \left( -i g_s \int_0^{\infty} d y^- A^+_\beta (y^- + w^-) T_\beta \right) \right]^\dagger  | X \rangle \times \\ \times \langle X | P \exp \left( -i g_s \int_0^{\infty} d y^{\prime -} A^+_\beta (y^{\prime -}) T_\beta \right) \fieldtensor^{+ j}_\alpha (0)  \; | P \rangle.
\end{multline}
The symbol $\sum_X$ includes all sums and integrals over final states.
Expanding the Wilson line in small coupling gives the relation,
\begin{equation}
\label{eq:WLexpand}
  P \exp \left( -i g_s \int_0^{\infty} d y^{-} A^+_\beta (y^{-}) T_\beta \right)
  = 1 + P \sum_{N=1}^\infty \frac{ (-i g_s )^N }{ N! }
    \prod_{i=1}^N \int_0^\infty d y_i^- A^+_\beta (y_i^{-}) T_\beta.
\end{equation}
The Feynman rules for the gluon PDF are found by
using Eqs.~(\ref{eq:WLexpand}), (\ref{eq:fieldtensor}) inside Eq.~(\ref{eq:gluedistb}) 
and directly applying the rules of ordinary perturbation theory.
\subsection{Lowest Order}
We now expand the operators in Eq.~(\ref{eq:gluedistb}) in powers of $g_s$.
At zeroth order, only the derivative terms in Eq.~(\ref{eq:fieldtensor})
contribute to Eq.~(\ref{eq:gluedistb}):
\begin{equation}
\label{eq:gluedistc}
\gluedist_{g/p}(\xi,P) = \sum_X \sum_{\alpha} \sum_{j=1}^2 \int \frac{dw^-}{2 \pi \xi P^+}    
e^{-i \xi P^+ w^- } \langle P | \,  (\partial^+ A_\alpha^j(w^-) - \partial^j A_\alpha^+(w^-))  \, | X \rangle \langle X | \, (\partial^+ A_\alpha^j(0) - \partial^j A_\alpha^+(0)) \, | P \rangle.
\end{equation}
Directly applying the rules of perturbation theory yields
\begin{equation}
\label{eq:glue_dist1g}
\gluedist^{(1g)}_{g/p}(\xi,P) = \xi P^+ \int \frac{d k^- d^2 {\bf k}_t}{(2 \pi)^4} 
\sum_\alpha \sum_{j=1}^2 \left( g^{j \mu_1} - \frac{k^j \njet^{\mu_1} }{k \cdot \njet} \right) 
\left( g^{j \mu_2} - \frac{k^j \njet^{\mu_2}}{k \cdot \njet} \right) 
\pdfamp^\alpha_{\mu_1}(P;k) \pdfamp^{\alpha\,\dagger}_{\mu_2}(P;k),
\end{equation}
where the factors $\pdfamp^\alpha_{\mu_1}$ and $\pdfamp^{\alpha\,\dagger}_{\mu_2}{}$
denote the parts of the incoming target bubble to the left and right
of the final-state cut, with their color indices.
Equation~(\ref{eq:glue_dist1g}) is exactly 
what we asserted to be the gluon-density factor,
Eq.~(\ref{eq:oneg_pdf}), for DIS with a single 
target gluon.  To express Eq.~(\ref{eq:glue_dist1g}) diagrammatically, 
we define the following Feynman rules:
There is a lower bubble representing the incoming target,
\begin{equation}
\picineq{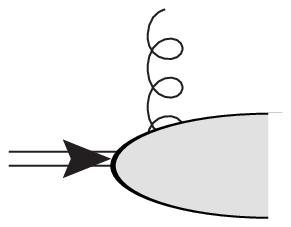}{0.6} = \pdfamp.
\end{equation}
For the $g^{j\mu}-k^j\njet^\mu/k\cdot\njet$ factors in
Eq.~(\ref{eq:glue_dist1g}) we use the notation
\begin{equation}
\label{eq:spec_vert1g}
\picineq{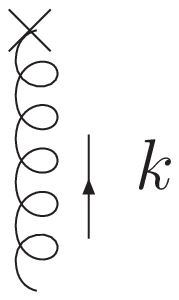}{0.6} = g^{j \mu_1} - \frac{k^j \njet^{\mu_1}}{k \cdot \njet}.
\end{equation}
Since Eq.~(\ref{eq:spec_vert1g}) modifies the vertex to the gluon from
the 
target bubble in Eq.~(\ref{eq:glue_dist1g}), we refer to it as the ``special'' vertex.
Diagrammatically, we notate the 
contribution to the gluon PDF from a single gluon emission on each side of the cut as
\begin{equation}
\gluedist^{(1g)}_{g/p}(\xi,P) = \picineq{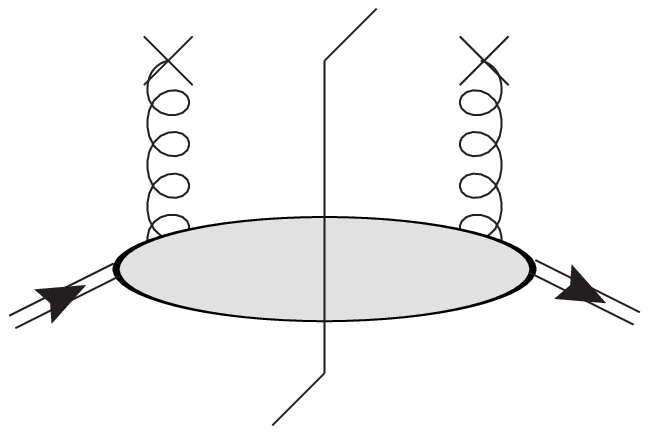}{0.6}.
\end{equation}
Implicit in this notation are an overall factor of $\xi P^+$, integrals
of $k^-$ and $k_t$, and a sum over the gluon color index $\alpha$.  Then
Eq.~(\ref{eq:fact_oneg}) can be written as
\begin{multline}
\label{eq:fact_onegb}
W \sim \int \frac{ d k^+ }{ 2k^+ }
 \sum_j \left( \picineq{ampsq_onea}{0.28} + \picineq{ampsq_oneb}{0.28} + \picineq{ampsq_onec}{0.28} + \picineq{ampsq_oned}{0.28}  \right) \times \\ \times
\left( \picineq{pdf_1g}{0.4} \right) - \text{subtraction terms} + \mathcal{O}(\Lambda/Q).
\end{multline}

\subsection{Beyond Lowest Order}
General Feynman rules for the gluon PDF, Eq.~(\ref{eq:gluedist}), are
found \cite{Collins:1981uw} 
by keeping other terms in the expansion of the Wilson line operator,
and the 
third term in Eq.~(\ref{eq:fieldtensor}).
To start out, let us ignore the third term in Eq.~(\ref{eq:fieldtensor}).
Looking only at the operators on one side of the final-state cut and using  
Eq.~(\ref{eq:WLexpand}), we get factors in the integrand of the form
\begin{equation}
\label{eq:Ngluonsa}
  -i \left( k_1 \cdot \njet g^{j \mu_1} - k_1^j \njet^{\mu_1} \right)
  \frac{ ig_s \njet^{\mu_2} f_{\alpha_1 \alpha_2 \kappa_2} }
       { \sum_{i=2}^N k_i \cdot \njet -i0 } 
  \times \frac{ ig_s \njet^{\mu_3} f_{\kappa_2 \alpha_3 \kappa_3} }
         { \sum_{i=3}^N k_i \cdot \njet -i0} \times \cdots \times 
  \frac{ ig_s \njet^{\mu_N} f_{\kappa_{N-1} \alpha_N \kappa_N} }
       { k_N \cdot \njet -i0 } 
  \pdfamp^{\alpha_1 \cdots \alpha_N}_{\mu_1 \cdots \mu_N}.
\end{equation}
At this order of perturbation theory, $N$ gluons are emitted from the
target bubble,
with color indices $\alpha_1, \cdots, \alpha_N$.

Now we treat the contributions
which use the third term in the field-strength tensor
Eq.~(\ref{eq:fieldtensor}). 
To stay at the same order of perturbation theory as in (\ref{eq:Ngluonsa}), we must drop down one order in the 
expansion of the Wilson line operator.  Thus, we obtain contributions of the form
\begin{equation}
\label{eq:Ngluonsb}
  g_sf_{\alpha_1 \alpha_2 \kappa_2} g^{j \mu_1}\njet^{\mu_2} 
  \times \frac{ ig_s \njet^{\mu_3} f_{\kappa_2 \alpha_3 \kappa_3} }
         { \sum_{i=3}^N k_i \cdot \njet -i0} \times \cdots \times 
  \frac{ ig_s \njet^{\mu_N} f_{\kappa_{N-1} \alpha_N \kappa_N} }
       { k_N \cdot \njet -i0 } 
  \pdfamp^{\alpha_1 \cdots \alpha_N}_{\mu_1 \cdots \mu_N}.
\end{equation}
Rather than treating the derivative terms and the third term 
in Eq.~(\ref{eq:gluedist}) separately, we can combine (\ref{eq:Ngluonsa}) and (\ref{eq:Ngluonsb}) 
at each order of perturbation theory.  The complete contribution is then
\begin{equation}
\label{eq:Ngluonsc}
  -i \left( k \cdot \njet g^{j \mu_1} - k_1^j \njet^{\mu_1} \right)
  \frac{ ig_s \njet^{\mu_2} f_{\alpha_1 \alpha_2 \kappa_2} }
       { \sum_{i=2}^N k_i \cdot \njet -i0 } 
  \times \frac{ ig_s \njet^{\mu_3} f_{\kappa_2 \alpha_3 \kappa_3} }
         { \sum_{i=3}^N k_i \cdot \njet -i0} \times \cdots \times 
  \frac{ ig_s \njet^{\mu_N} f_{\kappa_{N-1} \alpha_N \kappa_N} }
       { k_N \cdot \njet -i0 } 
  \pdfamp^{\alpha_1 \cdots \alpha_N}_{\mu_1 \cdots \mu_N}.
\end{equation}
The first factor in (\ref{eq:Ngluonsc}) tells us what the special 
gluon vertex analogous to Eq.~(\ref{eq:spec_vert1g}) is for $N$
gluons.
It differs from the corresponding factor in (\ref{eq:Ngluonsa}) by
having the gluon momentum $k_1$ replaced by the total momentum
$k$ in the first term only. 
We therefore obtain the Feynman rules for the gluon
PDF shown in Fig.~\ref{fig:feynman_rules}, where factors $-i$ and $i$
on the left and the right of the final-state cut have been dropped,
and we have removed a factor of $k^+=k\cdot\njet$, just as in
Eq.~(\ref{eq:spec_vert1g}). 
A general contribution to the gluon PDF is illustrated in Fig~\ref{fig:Fey_rul_sample}.
To simplify expressions, in later sections we will drop the explicit appearance of $-i0$ 
in the eikonal denominators. 
\begin{figure*}
\centering
\begin{tabular}{ccc}
    \includegraphics[scale=0.5]{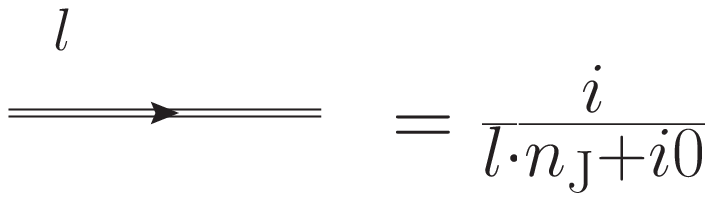} \hspace{12mm}  
&    \includegraphics[scale=0.5]{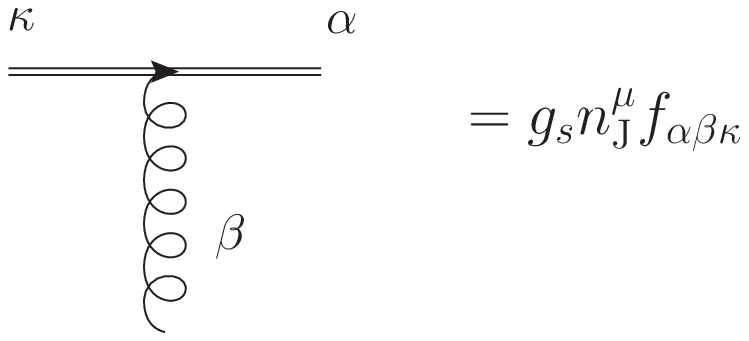} \hspace{12mm}
&    \includegraphics[scale=0.5]{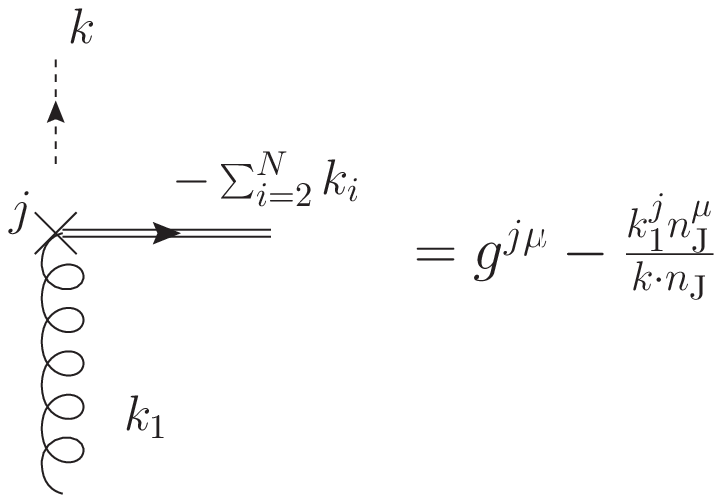}
\\
(a) & (b) & (c)
  \end{tabular}
\caption{Feynman rules for the gluon distribution function.}
\label{fig:feynman_rules}
\end{figure*}
\begin{figure*}
    \includegraphics[scale=0.4]{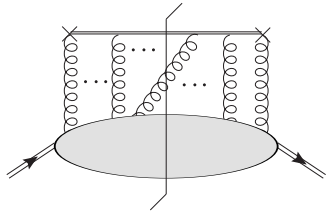}
  \caption{A general contribution to the gluon distribution function.}
  \label{fig:Fey_rul_sample}
\end{figure*}

\section{Scattering Off Two Collinear Gluons}
\label{sec:2gluons}

We now extend the treatment of Sect.~\ref{sec:one_gluon} to the case of two
target-collinear gluons.  The graphs under consideration are shown in
Fig.~\ref{fig:amp_oneco}.
Our aim is to show that when the graphs are summed, Ward identities similar to what were used 
in Sect.~\ref{sec:one_gluon} lead to a factorized structure. 
For a consistent factorization formula, the resulting gluon PDF should correspond 
to what is obtained using the Feynman rules of Fig.~\ref{fig:feynman_rules}.

\begin{figure*}
  \centering
  \begin{tabular}{c@{\hspace*{3mm}}c@{\hspace*{3mm}}c}
    \includegraphics[scale=0.4]{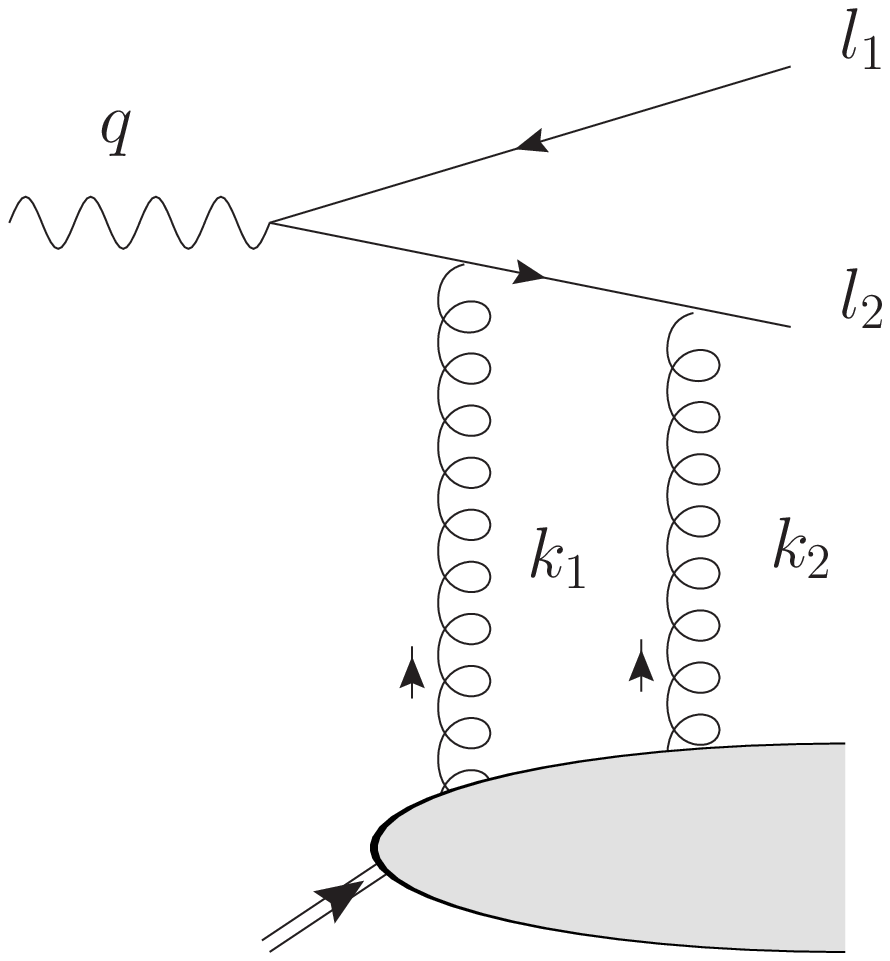}
    &
    \includegraphics[scale=0.4]{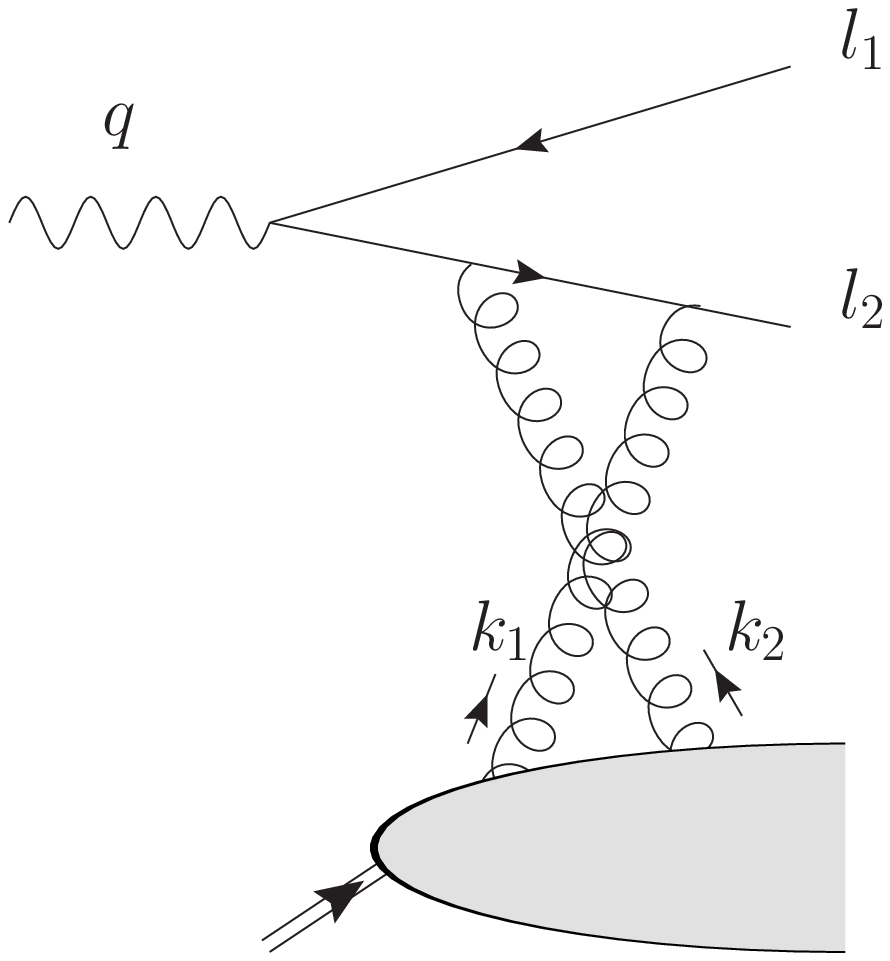}
    &
    \includegraphics[scale=0.4]{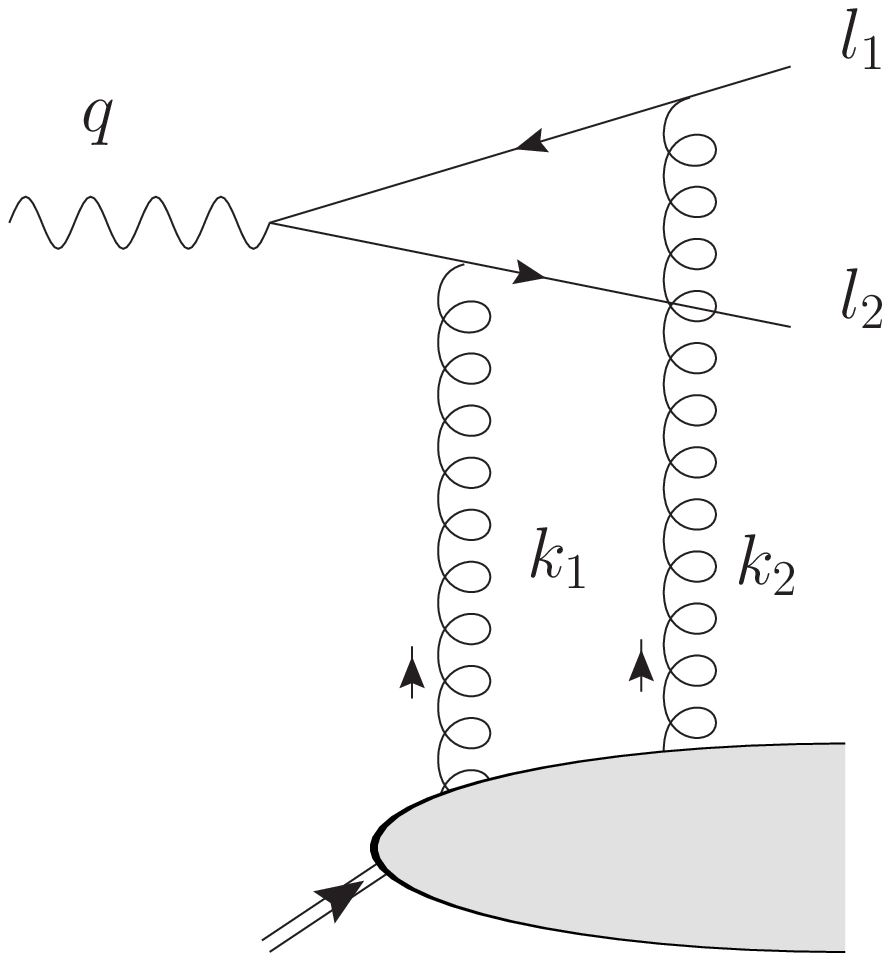}
  \\
    (a) & (b) & (c)
  \\[4mm]
    \includegraphics[scale=0.4]{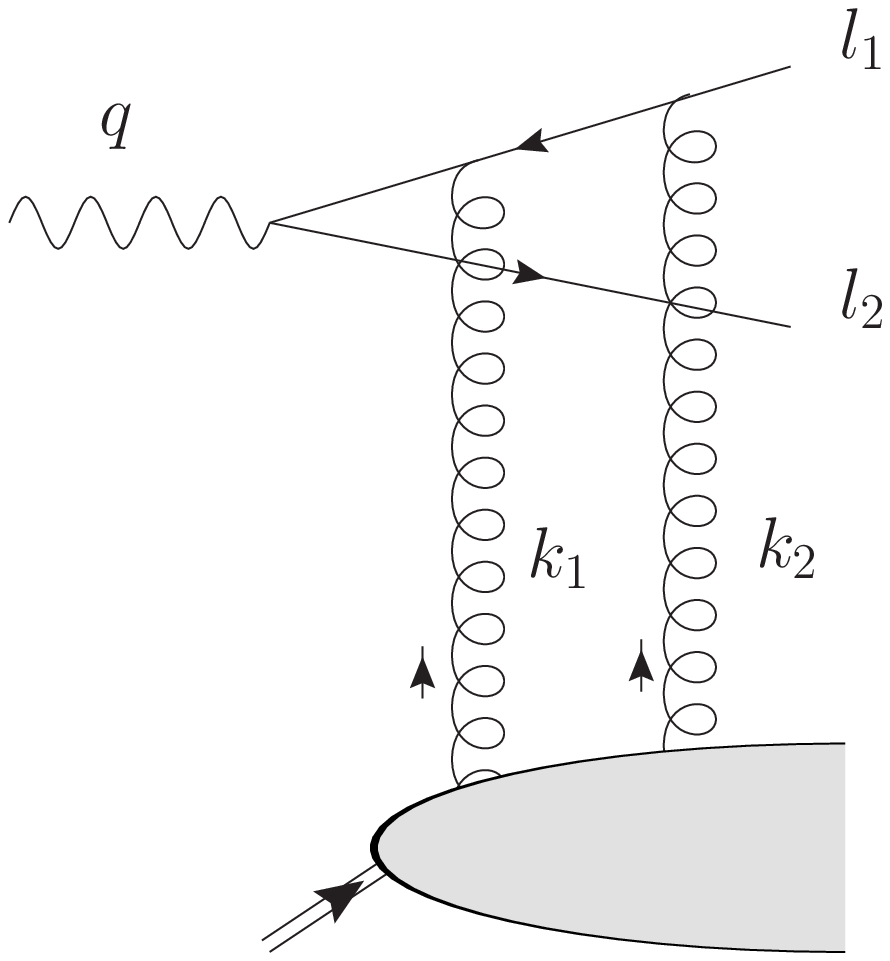}
    &
    \includegraphics[scale=0.4]{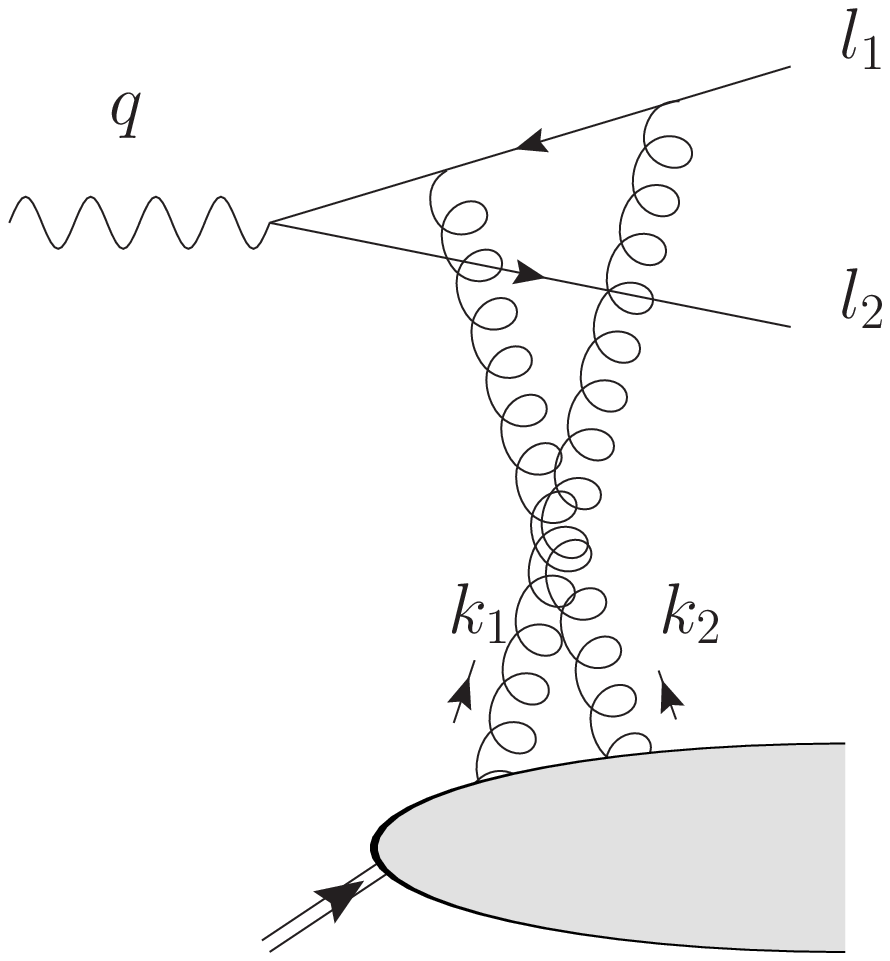}
    &
    \includegraphics[scale=0.4]{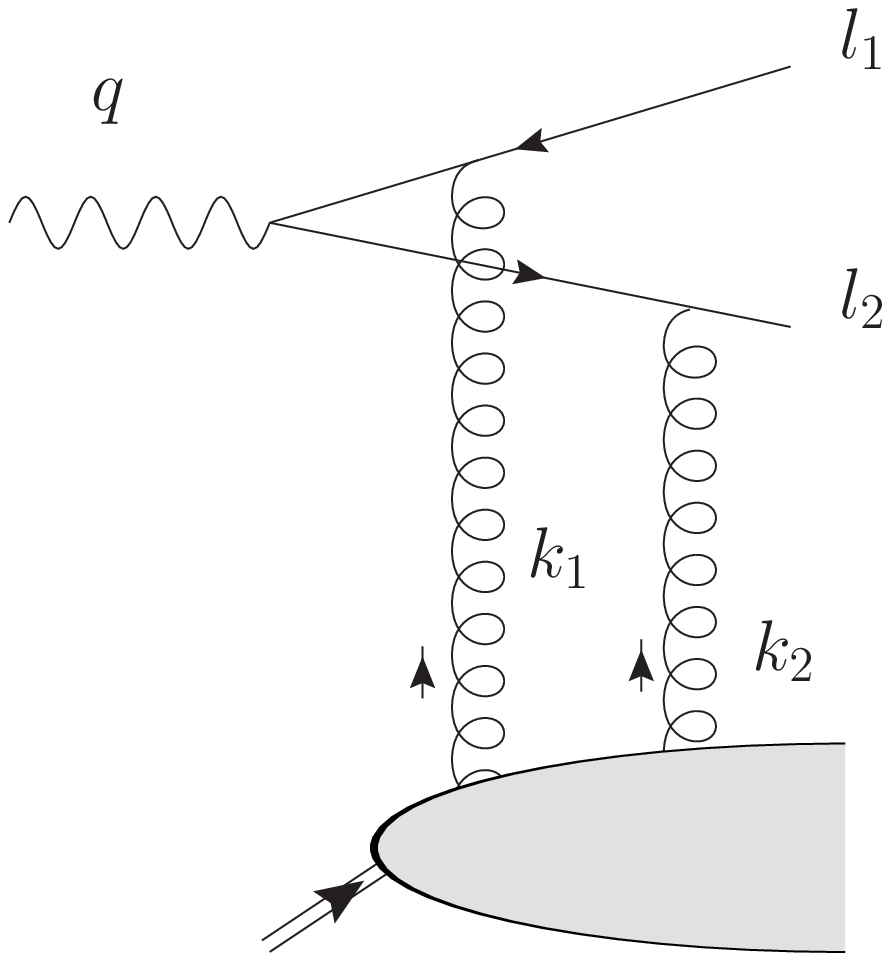}
  \\
    (d) & (e) & (f) 
  \end{tabular}
  \caption{DIS with two target-collinear gluons.}
  \label{fig:amp_oneco}
\end{figure*}

\subsection{Power counting and $K$, $G$ decomposition}

As in the previous sections, we write the amplitude obtained from the graphs in Fig.~\ref{fig:amp_oneco} as the contraction 
of an upper part and a lower part, connected by gluon-propagator
numerators: 
\begin{equation}
\label{eq:contraction}
M_{(2g)} =\upperbub_{\mu_1^{\prime} \mu_2^{\prime}} g^{\mu_1^{\prime} \mu_1} g^{\mu_2^{\prime} \mu_2} \pdfamp_{\mu_1 \mu_2}.
\end{equation}
All incoming gluons are target collinear, with the sizes of the momentum components given in Eq.~(\ref{eq:ksize}).
The gluons have leading contributions from regions where they are far off-shell, but these contribute only 
at high order in $g_s^2$.  Furthermore, since there are only gluon-fermion couplings in the upper part of the 
graph, ghosts do not enter at this level of the calculation.
A general, all-orders proof will need to take into account ghost attachments as well.

The upper part has one more external gluon than before, so its
dimension is reduced by a single power of $Q$; all the components of
$\upperbub_{\mu_1^{\prime} \mu_2^{\prime}}$ are therefore to be
treated as order $Q^{-1}$.  
The lower bubble now includes two Lorentz indices, and in the boosted frame 
has, 
up to overall powers of order $\Lambda$,
the power dependence,
\begin{align}
\label{eq:lowersize2g}
\pdfamp^{++} &\sim \frac{Q^2}{\Lambda^2}, \\
\pdfamp^{t+} \sim \pdfamp^{+ t} &\sim \frac{Q}{\Lambda}, \\
\pdfamp^{tt} \sim \pdfamp^{+-} \sim \pdfamp^{-+} &\sim Q^0, \\
\pdfamp^{t-} \sim \pdfamp^{-t}  &\sim \frac{\Lambda}{Q}, \\
\label{eq:lowersize2ge}
\pdfamp^{--} &\sim \frac{\Lambda^2}{Q^2}.
\end{align}
Combined with the $Q^{-1}$ factor from $\upperbub$, this gives
a super-leading contribution from the $++$ components, leading
contributions from the $+t$ and $t+$ components, and power-suppressed
contributions from the other components.

That the $++$ term gives a superleading result is an example of a
general result from power-counting arguments: A superleading
contribution arises whenever all the gluons joining the collinear and
hard subgraphs are longitudinally polarized.  We will show that there
is a cancellation after a sum over all graphs for $\upperbub$ leaving
a result that is just leading power.  That there is a leading
contribution from the $t+$ and $+t$ terms is expected: They correspond
to a hard scattering on a transversely polarized gluon accompanied by
a longitudinally polarized gluon.  The remaining terms are of a
non-leading power, and we can therefore ignore them.  We will show
that after a sum over graphs, the total of the $++$, $+t$ and $t+$
terms is equivalent at leading power to a hard scattering off a single
transversely polarized gluon multiplied by the appropriate two-gluon
factor obtained from the Feynman rules for the gluon PDF.

Just as with the single-gluon case, we split the propagator numerators
into $K$ and $G$ terms:
\begin{align}
\label{eq:gramyena}
g^{\mu_1^{\prime} \mu_1} & = K_1^{\mu_1^{\prime} \mu_1} + G_1^{\mu_1^{\prime} \mu_1}, \\
\label{eq:gramyenb}
g^{\mu_2^{\prime} \mu_2} & = K_2^{\mu_2^{\prime} \mu_2} + G_2^{\mu_2^{\prime} \mu_2}, 
\end{align}
where $K_1$, $K_2$, $G_1$ and $G_2$ 
are defined just like $K$, $G$,
except for the replacement of $k$ by the appropriate gluon momentum,
$k_1$ or $k_2$.  So we write Eq.~(\ref{eq:contraction}) as
\begin{equation}
\label{eq:contraction2}
M_{(2g)} =\upperbub_{\mu_1^{\prime} \mu_2^{\prime}} 
\left(K_1^{\mu_1^{\prime} \mu_1} K_2^{\mu_2^{\prime} \mu_2} + G_1^{\mu_1^{\prime} \mu_1} K_2^{\mu_2^{\prime} \mu_2}  + 
K_1^{\mu_1^{\prime} \mu_1} G_2^{\mu_2^{\prime} \mu_2} + G_1^{\mu_1^{\prime} \mu_1} G_2^{\mu_2^{\prime} \mu_2}  \right) \pdfamp_{\mu_1 \mu_2}.
\end{equation}
Exactly as in the previous section, the $K_1K_2$ term gives a
super-leading contribution, graph-by-graph, while the $K_1G_2$ and
$G_1K_2$ terms are leading.  The $G_1G_2$ term is power suppressed,
and so we neglect it.  Each of the 
non-suppressed terms has a factor of $K_1$
or $K_2$, so we will next apply gauge invariance to get a result of the
general form
\begin{equation}
\label{eq:amp_fact_2g}
M_{(2g)} = \upperbub_{\rho^{\prime}}(k_1 + k_2) \; g^{\rho^{\prime} \rho} \left( \cdots \right)_\rho + \mathcal{O}\left( \Lambda/Q \right).
\end{equation}
The first factor is just the basic amplitude already considered in the
previous section, but with gluon momentum $k_1+k_2$:
\begin{equation}
\label{eq:upperk1k2}
\upperbub_{\rho^{\prime}}(k_1 + k_2) = 
\left( \picineq{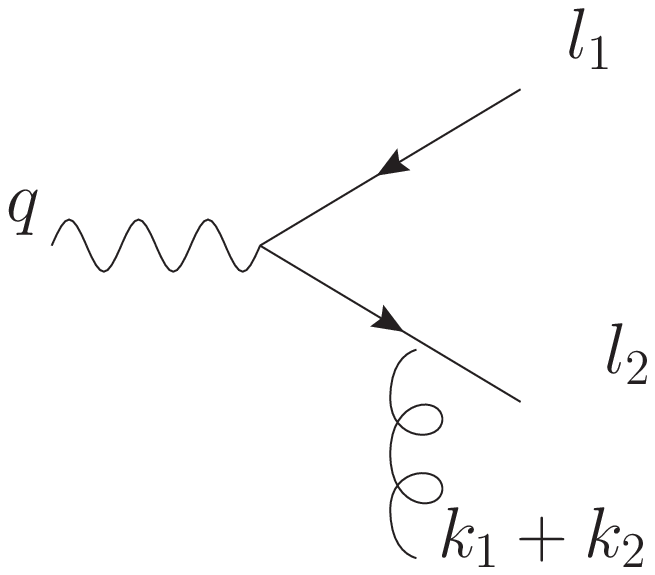}{0.3}  + \picineq{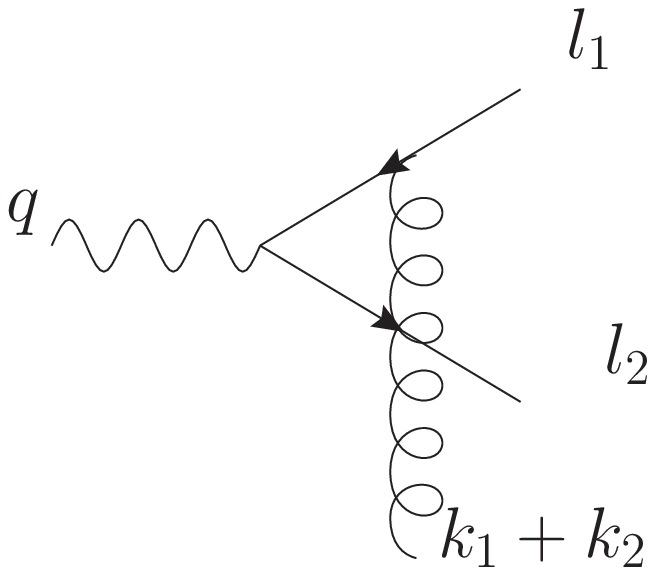}{0.3} \right)_{\rho^{\prime}}.
\end{equation}
The remaining factor $\left( \cdots \right)_{\rho}$ will determine what
``special''
vertex, analogous to Eq.~(\ref{eq:spec_vert1g}), we should use for the gluon density.

\subsection{Gauge invariance on $K_2$ factor}

We first apply the method of Sec.\ \ref{sec:WI.basic} to a single
$K$-gluon in Eq.\ (\ref{eq:contraction2}). 
The explicit expression for the upper bubbles 
in the first three graphs in Fig.~\ref{fig:amp_oneco} is
\begin{multline}
\label{eq:sumgraphs}
\upperbub^{\mu_1^{\prime} \mu_2^{\prime}}_{\text{(a--c)}} 
= g_s^2 \bar{u}(l_2) \left[ \gamma^{\mu_2^{\prime}} \left( \frac{1}{\slashed{l}_2 - \slashed{k}_2 - m} \right) \gamma^{\mu_1^{\prime}} 
\left(  \frac{1}{\slashed{l}_2 - \slashed{k}_1 - \slashed{k}_2 - m} \right) \gamma^{\nu} t_\beta t_\alpha + \right. \\ +
\gamma^{\mu_1^{\prime}} \left( \frac{1}{\slashed{l}_2 - \slashed{k}_1 - m} \right) \gamma^{\mu_2^\prime}
\left(  \frac{1}{\slashed{l}_2 - \slashed{k}_1 - \slashed{k}_2 - m} \right) \gamma^{\nu} t_\alpha t_\beta + 
\left. \gamma^{\mu_1^{\prime}} \left( \frac{1}{\slashed{l}_2 - \slashed{k}_1 - m} \right) \gamma^{\nu}
\left(  \frac{1}{\slashed{k}_2 - \slashed{l}_1 - m} \right) \gamma^{\mu_2^\prime} t_\alpha t_\beta \right] v(l_1).
\end{multline}
We organize these graphs by having the $k_1$ gluon attached in one place
on the quark line, and then the graphs correspond to the different
possible placements of the $k_2$ gluon.   Similarly, for the other
three graphs,
\begin{multline}
\label{eq:sumgraphsb}
\upperbub^{\mu_1^{\prime} \mu_2^{\prime}}_{\text{(d--f)}} 
= g_s^2 \bar{u}(l_2) \left[ \gamma^{\nu} \left( \frac{1}{\slashed{k}_2 + \slashed{k}_1 - \slashed{l}_1 - m} \right) \gamma^{\mu_1^{\prime}} 
\left(  \frac{1}{\slashed{k}_2 - \slashed{l}_1 - m} \right) \gamma^{\mu_2^{\prime}} t_\alpha t_\beta + \right. \\ +
\gamma^{\nu} \left( \frac{1}{\slashed{k}_1 + \slashed{k}_2 - \slashed{l}_1 - m} \right) \gamma^{\mu_2^\prime}
\left(  \frac{1}{\slashed{k}_1 - \slashed{l}_1 - m} \right) \gamma^{\mu_1^{\prime}} t_\beta t_\alpha + 
\left. 
\gamma^{\mu_2^{\prime}} \left( \frac{1}{\slashed{l}_2 - \slashed{k}_2 - m} \right) \gamma^{\nu}
\left(  \frac{1}{\slashed{k}_1 - \slashed{l}_1 - m} \right) \gamma^{\mu_1^{\prime}} t_\beta t_\alpha
 \right] v(l_1).
\end{multline}

In Eq.~(\ref{eq:contraction2}), the terms that are leading 
or super-leading
have at
least one $K$-gluon, to which we apply the method of Sec.\
\ref{sec:WI.basic}.  We start with the case of any term involving
$K_2$.  For graphs (a)--(c), we have a factor
\begin{multline}
\left[ \upperbub_{\mu_1^{\prime} \mu_2^{\prime}} K_2^{\mu_2^{\prime} \mu_2} \right]_{\text{(a--c)}} = g_s^2 \bar{u}(l_2) \left[ \slashed{k}_2 \left( \frac{1}{\slashed{l}_2 - \slashed{k}_2 - m} \right) \gamma_{\mu_1^{\prime}}
\left(  \frac{1}{\slashed{l}_2 - \slashed{k}_1 - \slashed{k}_2 - m} \right) \gamma^{\nu} t_\beta t_\alpha  + \right. \\ +
\gamma_{\mu_1^{\prime}} \left( \frac{1}{\slashed{l}_2 - \slashed{k}_1 - m} \right) \slashed{k}_2
\left(  \frac{1}{\slashed{l}_2 - \slashed{k}_1 - \slashed{k}_2 - m} \right) \gamma^{\nu} t_\alpha t_\beta  + 
\left. \gamma_{\mu_1^{\prime}} \left( \frac{1}{\slashed{l}_2 - \slashed{k}_1 - m} \right) \gamma^{\nu}
\left(  \frac{1}{\slashed{k}_2 - \slashed{l}_1 - m} \right) \slashed{k}_2 t_\alpha t_\beta  \right] \frac{\njet^{\mu_2}}{k_2 \cdot \njet} v(l_1).
\end{multline}
Calling the terms in brackets term 1, term 2, and term 3, we now make 
use of the following identities:
\begin{align}
\slashed{k}_2 &= -(\slashed{l}_2 - \slashed{k}_2 - m) + (\slashed{l}_2 - m) && \text{in term 1}, \\
&= -(\slashed{l}_2 - \slashed{k}_1 - \slashed{k}_2 - m) + (\slashed{l}_2 - \slashed{k}_1 - m)  &&\text{in term 2}, \\
&= (\slashed{k}_2 - \slashed{l}_1 - m) + (\slashed{l}_1 + m)  &&\text{in term 3}.
\end{align} 
Using the Dirac equation to eliminate two of the terms, we then have
\begin{multline}
\left[ \upperbub_{\mu_1^{\prime} \mu_2^{\prime}} K_2^{\mu_2^{\prime} \mu_2} \right]_{\text{(a--c)}} 
= g_s^2 \bar{u}(l_2) \left[ - \gamma_{\mu_1^{\prime}}
\left(  \frac{1}{\slashed{l}_2 - \slashed{k}_1 - \slashed{k}_2 - m} \right) \gamma^{\nu} t_\beta t_\alpha  + \right. \\ +
\gamma_{\mu_1^{\prime}} 
\left(  \frac{1}{\slashed{l}_2 - \slashed{k}_1 - \slashed{k}_2 - m} \right) \gamma^{\nu} t_\alpha t_\beta  - 
\gamma_{\mu_1^{\prime}} 
\left(  \frac{1}{\slashed{l}_2 - \slashed{k}_1 - m} \right) \gamma^{\nu} t_\alpha t_\beta  + 
\left. 
\gamma_{\mu_1^{\prime}} 
\left(  \frac{1}{\slashed{l}_2 - \slashed{k}_1 - m} \right) \gamma^{\nu} t_\alpha t_\beta 
\right] v(l_1) \frac{\njet^{\mu_2}}{k_2 \cdot \njet}.
\end{multline}
The last two terms cancel because of the Abelian nature of the QED coupling.  However, the remaining terms lead to 
a non-vanishing result due to the non-vanishing commutation relations for 
$t_\alpha$ and $t_\beta$,
\begin{equation}
\label{eq:K2contracta}
\left[ \upperbub_{\mu_1^{\prime} \mu_2^{\prime}} K_2^{\mu_2^{\prime} \mu_2} \right]_{\text{(a--c)}} 
= g_s \bar{u}(l_2)  \gamma_{\mu_1^{\prime}} t_{\kappa}
\left(  \frac{1}{\slashed{l}_2 - \slashed{k}_1 - \slashed{k}_2 - m} \right) \gamma^{\nu} 
 v(l_1) \; \left[ ig_s f_{\alpha \beta \kappa} \frac{\njet^{\mu_2}}{k_2 \cdot \njet} \right].
\end{equation}
Similarly for graphs (d)--(f) we have
\begin{equation}
\label{eq:K2contractb}
\left[ \upperbub_{\mu_1^{\prime}  \mu_2^{\prime}} K_2^{\mu_2^{\prime} \mu_2} \right]_{\text{(d--f)}} 
= g_s \bar{u}(l_2)  \gamma^{\nu} 
\left(  \frac{1}{\slashed{k}_2 + \slashed{k}_1 - \slashed{l}_1 - m} \right) \gamma_{\mu_1^{\prime}} t_{\kappa} 
 v(l_1) \; \left[ i g_s f_{\alpha \beta \kappa} \frac{\njet^{\mu_2}}{k_2 \cdot \njet} \right].
\end{equation}
Adding Eqs.~(\ref{eq:K2contracta}) and (\ref{eq:K2contractb}),  we get
\begin{equation}
\label{eq:K2contraction}
\upperbub_{\mu_1^{\prime}  \mu_2^{\prime}} K_2^{\mu_2^{\prime} \mu_2} 
= g_s \bar{u}(l_2)  
\left[ 
    \gamma_{\mu_1^{\prime}} t_{\kappa}
    \left(  \frac{1}{\slashed{l}_2 - \slashed{k}_1 - \slashed{k}_2 - m}
    \right) \gamma^{\nu}  
\right. + \left. 
      \gamma^{\nu} \left(  \frac{1}{\slashed{k}_2 + \slashed{k}_1 -
                \slashed{l}_1 - m} \right) \gamma_{\mu_1^{\prime}} t_{\kappa} 
\right]
 v(l_1) \left[ ig_s f_{\alpha \beta \kappa} \frac{\njet^{\mu_2}}{k_2 \cdot \njet} \right].
\end{equation}
This is the general expression for the upper bubble when 
it is contracted with the $K_2$-term from gluon 2.  Notice that it
vanishes when the theory is Abelian.  This is an example of a general
result that $K$-gluons give 
zero contribution in an Abelian gauge
theory. 
Notice also that it is of the form of the upper factor
(\ref{eq:one.gluon}) for one gluon of momentum $k_1+k_2$, times an
eikonal factor.

\subsection{Gauge invariance on $K_1$ factor}

If we instead consider the contraction of the 
upper bubble with the $K_1$ gluon from gluon 1, then 
exactly similar steps lead to,
\begin{equation}
\label{eq:K1contraction}
\upperbub_{\mu_1^{\prime} \mu_2^{\prime}} K_1^{\mu_1^{\prime} \mu_1} 
= -g_s \bar{u}(l_2)  
\left[ 
    \gamma_{\mu_2^{\prime}} t_{\kappa}
    \left(  \frac{1}{\slashed{l}_2 - \slashed{k}_1 - \slashed{k}_2 - m}
   \right) \gamma^{\nu}  
\right. +  \left. 
      \gamma^{\nu} \left(  \frac{1}{\slashed{k}_2 + \slashed{k}_1 -
          \slashed{l}_1 - m} \right) \gamma_{\mu_2^{\prime}} t_{\kappa} 
\right]
 v(l_1) \left[ ig_s f_{\alpha \beta \kappa} \frac{\njet^{\mu_1}}{k_1 \cdot \njet} \right].
\end{equation}
Note the overall minus sign that arises from the reversed roles
of gluon 1 and gluon 2.
To complete the analysis, we must consider the separate cases: where the other gluon is also 
a $K$-gluon, or where one gluon is a $K$-gluon and the other is a
$G$-gluon.

\subsection{$K_1K_2$ term}
\label{sec:K1K2}
The $K_1K_2$ term in Eq.~(\ref{eq:contraction2}) follows
immediately from Eq.~(\ref{eq:K2contraction}) when we include the
contraction with $K_1$:
\begin{multline}
\label{eq:factKK}
\upperbub_{\mu_1^{\prime} \mu_2^{\prime}} K_1^{\mu_1^{\prime} \mu_1} K_2^{\mu_2^{\prime} \mu_2} 
= g_s \bar{u}(l_2)  
\left[ 
   \gamma_{\rho} t_{\kappa} \left(  \frac{1}{\slashed{l}_2 - \slashed{k}_1 -
     \slashed{k}_2 - m} \right) \gamma^{\nu}  
+
    \gamma^{\nu} \left(  \frac{1}{\slashed{k}_2 + \slashed{k}_1 - \slashed{l}_1 -
    m} \right) \gamma_{\rho} t_{\kappa}  
\right] v(l_1) 
\times \\ \times 
 \left[ ig_s f_{\alpha \beta \kappa} k_{1}^{\rho} \frac{\njet^{\mu_1} \njet^{\mu_2}}{(k_1 \cdot \njet)(k_2 \cdot \njet)} \right].
\end{multline}
This has the form of the amplitude (\ref{eq:one.gluon}) with a single
gluon of momentum $k_1+k_2$ multiplied by a special vertex.  Thus for
the $K_1K_2$ term in Eq.~(\ref{eq:contraction2}), we have:
\begin{equation}
\label{eq:factKKb}
\upperbub^{\alpha\beta}_{\mu_1^{\prime} \mu_2^{\prime}} K_1^{\mu_1^{\prime} \mu_1} K_2^{\mu_2^{\prime} \mu_2} \pdfamp^{\alpha\beta}_{\mu_1 \mu_2}
= \upperbub^\kappa_{\rho^{\prime}}(k_1 + k_2) ~ g^{\rho^{\prime} \rho} 
\left[ i g_s f_{\alpha \beta \kappa} k_{1,\rho} \frac{\njet^{\mu_1} \njet^{\mu_2}}{(k_1 \cdot \njet)(k_2 \cdot \njet)} \right] \pdfamp^{\alpha\beta}_{\mu_1 \mu_2}.
\end{equation}
The two $K$-gluons have been factored out into eikonal couplings.  A diagrammatic 
representation of Eq.~(\ref{eq:factKK}) is shown in
Fig.~\ref{fig:k1k2_fact}.  
Note that this term would exactly vanish in an Abelian theory, where $f_{\alpha \beta \kappa} = 0$.
\begin{figure}
\centering
    \includegraphics[scale=0.4]{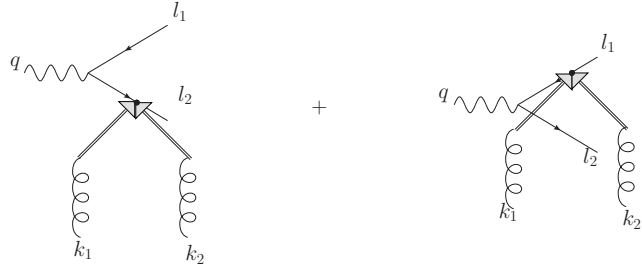}
\caption{Graphical structure of the $K_1$-$K_2$ term.  The arrows are used to represent $K$-gluons which couple 
to the quark via the special coupling in Eq.~(\ref{eq:factKKb}).}
\label{fig:k1k2_fact}
\end{figure}

Graph-by-graph, this is still super-leading, since the factor of
$g^{\rho'\rho}$ and everything to its right power-counts just like
$\pdfamp^{\rho'}$ for the case of one-gluon exchange.  Furthermore,
(\ref{eq:factKKb}) is not symmetric between the gluon momenta, $k_1$
and $k_2$, despite the symmetric occurrence of $K_1$ and $K_2$ factors
on the left-hand side.  Both problems are remedied by applying the
$K+G$ decomposition to the $g^{\rho^{\prime}\rho}$ factor in
Eq.~(\ref{eq:factKKb}):
\begin{equation}
\label{eq:gramyenc}
g^{\rho^{\prime} \rho} =  K_{(1+2)}^{\rho^{\prime} \rho} + G_{(1+2)}^{\rho^{\prime} \rho},  
\end{equation}
where
\begin{align}
K_{(1+2)}^{\rho^{\prime} \rho} & =  \frac{(k_1+k_2)^{\rho^{\prime}}}{(k_1 + k_2) \cdot \njet} \njet^{\rho},
\\
G_{(1+2)}^{\rho^{\prime} \rho} & =   g^{\rho^{\prime} \rho} 
- \frac{(k_1 + k_2)^{\rho^{\prime}}}{(k_1 + k_2) \cdot \njet}  \njet^{\rho}.
\end{align}
The $K_{(1+2)}$ term gives zero when multiplied with
$\upperbub_{\rho^{\prime}}(k_1 + k_2)$ (see the calculation in Sect.~\ref{sec:WI.basic}).  This removes the super-leading part,
just as in one-gluon exchange, leaving only the $G_{(1+2)}$ term.
Hence,
the $K_1K_2$ term is 
\begin{multline}
\label{eq:factKKG}
\upperbub_{\rho^{\prime}}(k_1 + k_2) G_{(1+2)}^{\rho'\rho}
\left[ i g_s f_{\alpha \beta \kappa} k_{1,\rho} \frac{\njet^{\mu_1} \njet^{\mu_2}}{(k_2 \cdot
    \njet)(k_2 \cdot \njet)} 
\right] \pdfamp^{\alpha\beta}_{\mu_1 \mu_2} 
\\
=
\upperbub_{\rho^{\prime}}(k_1 + k_2)
\frac{ i g_s f_{\alpha \beta \kappa} \njet^{\mu_1} \njet^{\mu_2} }
     { (k_1 \cdot \njet)(k_2 \cdot \njet)(k_1 + k_2) \cdot \njet }
     \left( k_1^{\rho^{\prime}} (k_2 \cdot \njet) - k_2^{\rho^{\prime}} (k_1 \cdot \njet) \right) 
\pdfamp^{\alpha\beta}_{\mu_1 \mu_2}.
\end{multline}
Thus it has the structure of Eq.~(\ref{eq:amp_fact_2g}).

\subsection{$G_1K_2$ term}
We next apply Eq.~(\ref{eq:K2contraction}) to the $G_1K_2$ term to
obtain
\begin{equation}
\label{eq:factG1K2b}
\upperbub^{\alpha\beta}_{\mu_1^{\prime} \; \mu_2^{\prime}} G_1^{\mu_1^{\prime} \mu_1} K_2^{\mu_2^{\prime} \mu_2} \pdfamp^{\alpha\beta}_{\mu_1 \mu_2}
=  \upperbub^\kappa_{\rho}(k_1 + k_2) \left[ i g_s f_{\alpha \beta \kappa} \left(g^{\rho\mu_1} - \frac{k_1^\rho \njet^{\mu_1} }{k_1 \cdot \njet}  \right)\frac{\njet^{\mu_2}}{k_2 \cdot \njet} \right] \pdfamp^{\alpha\beta}_{\mu_1 \mu_2},
\end{equation}
again with the structure of Eq.~(\ref{eq:amp_fact_2g}).
As already observed, there is no super-leading contribution from the
$G_1K_2$ term, only a leading-power contribution.

\subsection{$K_1G_2$ term} 
For the $K_1G_2$ term we similarly obtain
\begin{equation}
\label{eq:factG2K1}
\upperbub^{\alpha\beta}_{\mu_1^{\prime} \; \mu_2^{\prime}} K_1^{\mu_1^{\prime} \mu_1} G_2^{\mu_2^{\prime} \mu_2} \pdfamp^{\alpha\beta}_{\mu_1 \mu_2}
=  \upperbub^\kappa_{\rho}(k_1 + k_2) 
   \left[ - i g_s f_{\alpha \beta \kappa} \left(g^{\rho\mu_2} - \frac{k_2^\rho \njet^{\mu_2} }{k_2 \cdot \njet}  \right)\frac{\njet^{\mu_1}}{k_1 \cdot \njet} \right] \pdfamp^{\alpha\beta}_{\mu_1 \mu_2}.
\end{equation}
Note the overall minus sign relative to Eq.~(\ref{eq:factG1K2b}) that arises because of the 
reversed role of gluons 1 and 2.  
A graphical representation of the factorization in
Eqs.~(\ref{eq:factG1K2b}) and (\ref{eq:factG2K1}) 
is shown in Fig.~\ref{fig:GKfact}.

\begin{figure*}
  \centering
  \begin{tabular}{c}
    \includegraphics[scale=0.4]{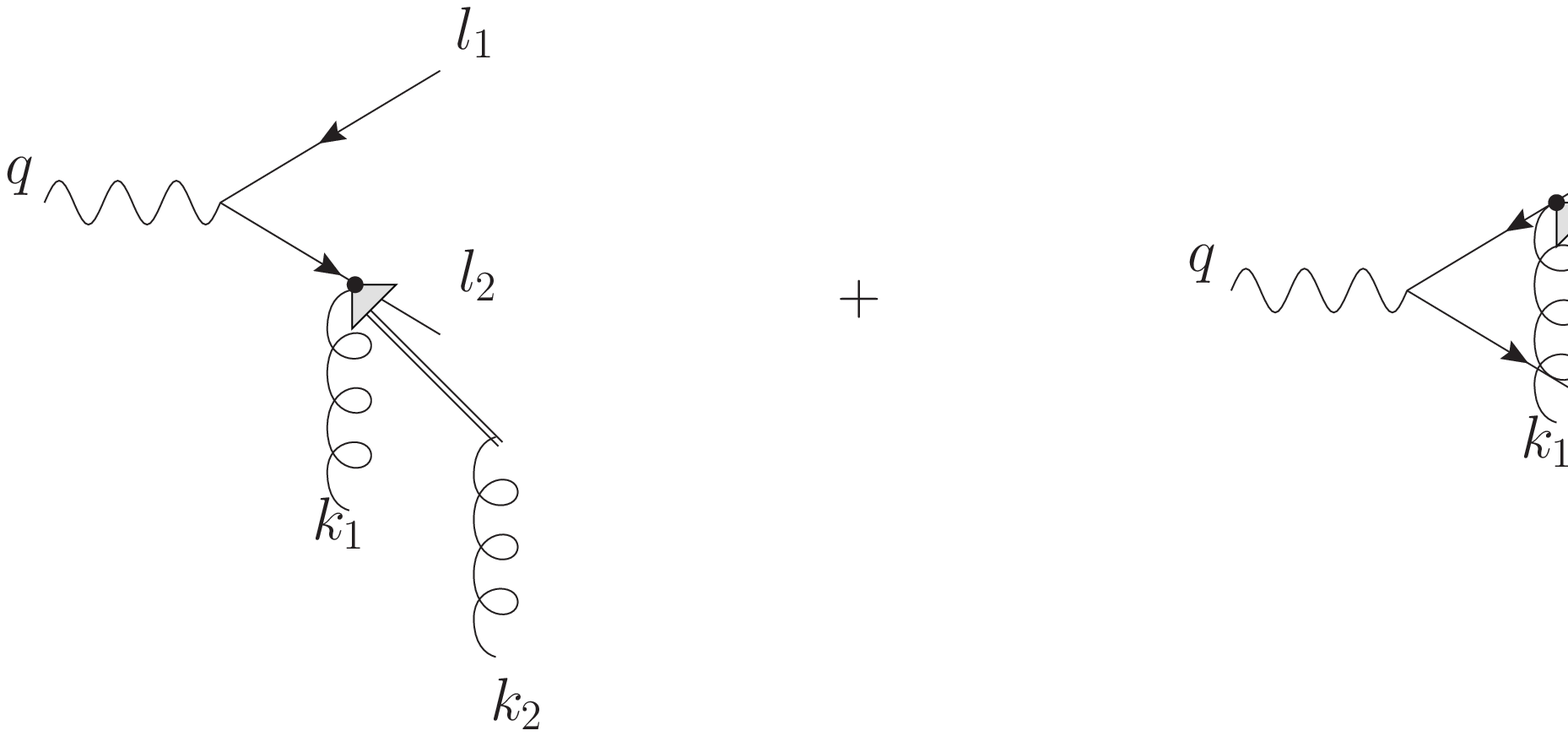}
  \\
    (a) 
  \\[4mm]
    \includegraphics[scale=0.4]{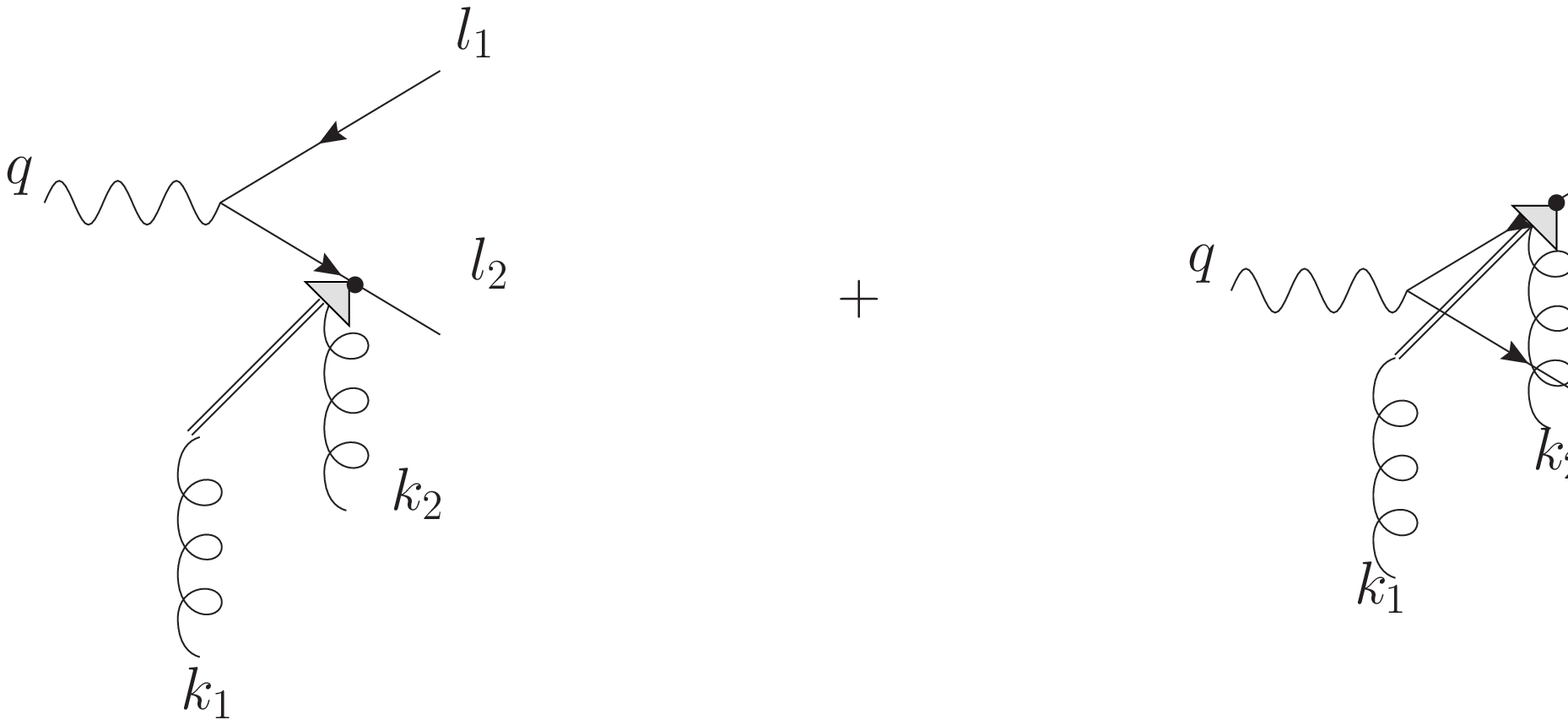}
  \\
    (b)
  \end{tabular}
  \caption{Graphical structure of (a) the $G_1$-$K_2$ term and (b) the $G_2$-$K_1$ term.}
  \label{fig:GKfact}
\end{figure*}

\subsection{Factorization and the Gluon Distribution Function}
Adding Eqs.~(\ref{eq:factKKG}), (\ref{eq:factG1K2b}), and
(\ref{eq:factG2K1}) gives
\begin{equation}
\label{eq:eikonal_fact}
M_{(2g)} =  
\upperbub^\kappa_\rho(k_1 + k_2) 
\left\{ ig_s f_{\alpha \beta \kappa} 
       \left[ \frac{g^{\rho\mu_{1}} \njet^{\mu_2}}{k_2 \cdot \njet}  - \frac{g^{\rho \mu_{2}} \njet^{\mu_1}}{k_1 \cdot \njet}  
              + \frac{\njet^{\mu_1} \njet^{\mu_2} }{(k_1 + k_2)\cdot\njet}
               \left( \frac{k_{2}^\rho}{k_1 \cdot \njet} - \frac{k_{1}^\rho}{k_2 \cdot \njet} \right)
        \right]
       \pdfamp^{\alpha\beta}_{\mu_1 \mu_2}
\right\}
+ \mathcal{O}\!\left( \Lambda/Q \right).
\end{equation}
This may be written compactly if we define the factor in braces to be
\begin{equation}
\label{eq:2gpdf}
\tilde{\pdfamp}^{\kappa,\rho}_{(2g)} =  
ig_s f_{\alpha \beta \kappa} \left[ \frac{g^{\rho\mu_{1}} \njet^{\mu_2}}{k_2 \cdot \njet}  - \frac{g^{\rho \mu_{2}} \njet^{\mu_1}}{k_1 \cdot \njet}  
+ \frac{\njet^{\mu_1} \njet^{\mu_2} }{(k_1 + k_2)\cdot \njet} \left( \frac{k_{2}^\rho}{k_1 \cdot \njet} - \frac{k_{1}^\rho}{k_2 \cdot \njet} \right) \right] \pdfamp^{\alpha\beta}_{\mu_1 \mu_2}.
\end{equation}
Normally a vector like this would have its $\rho=+$ component power
counting as $Q/\Lambda$.  In fact this component is 
zero.  The reason is that it is constructed from a combination of
$G_1$, $G_2$, and $G_{(1+2)}$, each of which gives this property
individually.

The importance of the last remark is that it shows that the dominant
contribution, i.e., the leading-power contribution, is from transverse
values of $\rho$, so that $\upperbub$, which goes into the
hard-scattering factor, 
can be treated as having an incoming transversely
polarized gluon. Compare this with what is found in 
Eq.~(\ref{eq:psdropa}).
  
It is the factor $\tilde{\pdfamp}^{\kappa,\rho}_{(2g)}$ which we would like to
identify with a two-gluon factor in the 
amplitude for the gluon distribution function. 
The derivation so far, where we have systematically eliminated
super-leading contributions, shows that we have a result with the
standard leading power of $Q$, i.e., $Q^0$.
The leading terms are for $\rho = j$, a transverse index:
\begin{equation}
\label{eq:2gleading}
\tilde{\pdfamp}^\kappa_{j\,(2g)}(P;k_1,k_2) = ig_s f_{\alpha \beta \kappa} 
   \left[ \frac{ \njet{}^{\mu_2} \pdfamp^{\alpha\beta}_{j\mu_2} }{ k_2 \cdot \njet} 
        - \frac{ \njet{}^{\mu_1} \pdfamp^{\alpha\beta}_{\mu_1 j} }{ k_1 \cdot \njet}  
        + \frac{ \njet{}^{\mu_1} \njet{}^{\mu_2} }{ (k_1 + k_2) \cdot \njet} 
          \left( \frac{k_{2\,j}}{k_1 \cdot \njet} - \frac{k_{1\,j}}{k_2 \cdot \njet} \right)
          \pdfamp^{\alpha\beta}_{\mu_1\mu_2}
    \right] \sim Q^0.
\end{equation}
The steps for obtaining a factorization formula are now exactly analogous 
to the steps in Sect.~\ref{sec:fact_steps}.
The analogue of Eq.~(\ref{eq:psdropa}) for the case of two gluons is
\begin{align}
\label{eq:twogluon_fact}
M_{(2g)}  &=  \sum_{j=1}^2 \upperbub^{j}(l_1,l_2;\hat{k})
\tilde{\pdfamp}_{(2g)}^{j}(P;k_1,k_2)
\nonumber\\
&=\sum_{j=1}^2 \left( \picineq{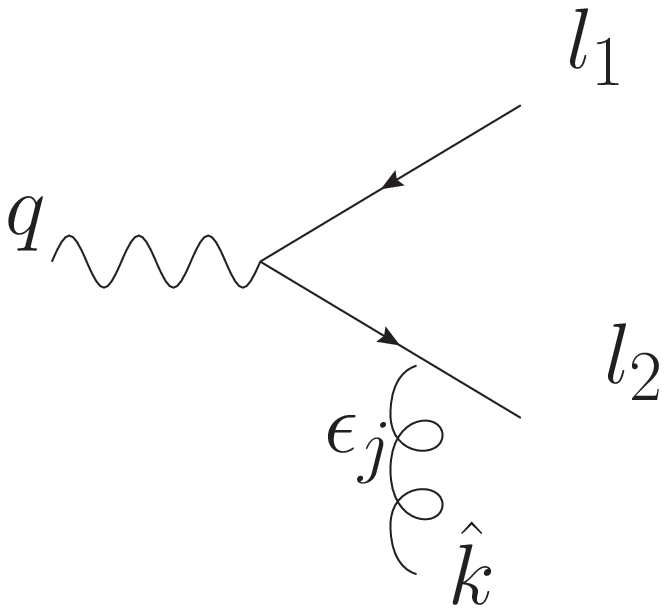}{0.3}  + \picineq{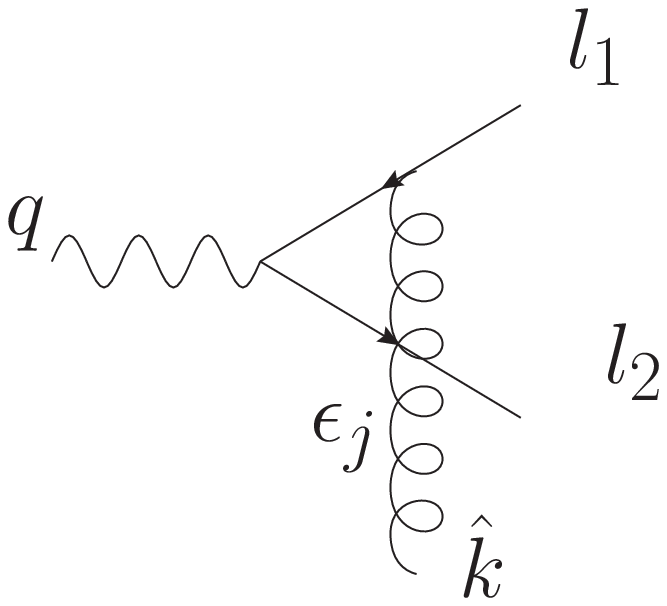}{0.3} \right)^{j} \left( \picineq{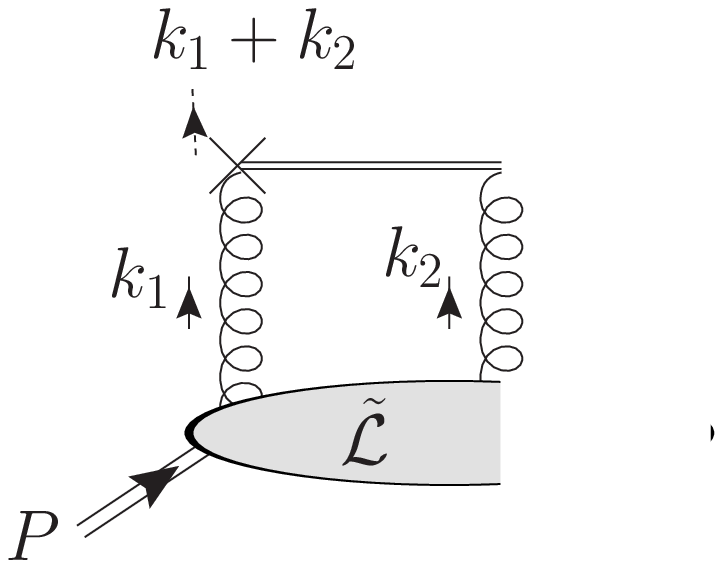}{0.35} + \picineq{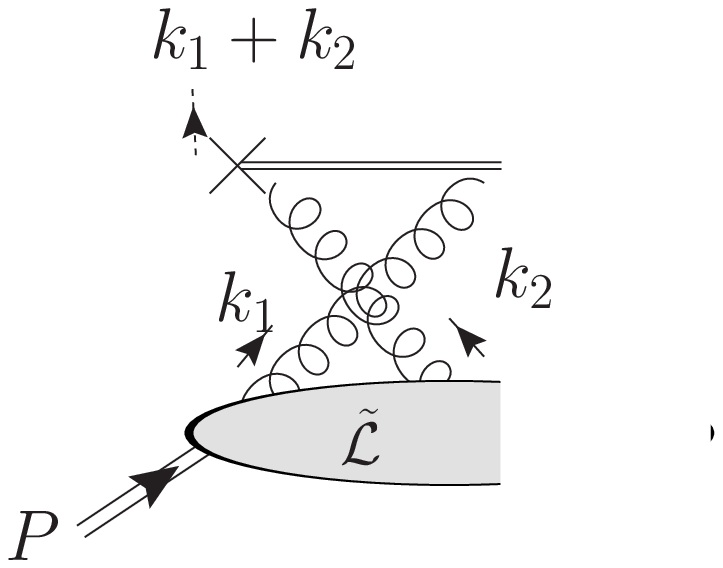}{0.35} \right)^{j} + \mathcal{O}(\Lambda/Q).
\end{align}
\end{widetext}
Here we have defined $k = k_1 + k_2$.

The factor $\tilde{\pdfamp}_{(2g)}^{j}(P;k_1,k_2)$, given in
Eq.~(\ref{eq:2gleading}), is exactly what is obtained
from the Feynman rules for the gluon PDF listed in
Fig.~\ref{fig:feynman_rules}.  This would not have happened if we had
neglected the $K_1 K_2$ contribution from Sect.~\ref{sec:K1K2} to the
leading behavior in Eq.~(\ref{eq:2gleading}).
Equation~(\ref{eq:factKKG}) is therefore important for agreement
between the Feynman rules for the gluon density and the leading
contribution to DIS.

To summarize this section, we find that leading-power contributions
arise when one or both gluons is longitudinally polarized.  With one
longitudinally polarized gluon, we get expected eikonal factors in
Eqs.~(\ref{eq:factG1K2b},\ref{eq:factG2K1}), after a sum over graphs.
These factors are analogous to the eikonal
propagators that appear in the quark PDF, but they do not have the
correct Feynman rules to correspond to the gluon PDF.  A further
contribution Eq.~(\ref{eq:factKKG}) arises when both gluons are
longitudinally polarized.  Previously, this contribution was expected
to vanish (see, e.g.,~\cite{Labastida:1984gy}).  But it is in fact
non-vanishing, and is needed for a correct correspondence with the
definition of the gluon PDF.

\section{Summary and Conclusion}
\label{sec:conclusion}
We have given a direct illustration using gluon-induced DIS
that contributions from longitudinally polarized gluons do not cancel in sums over graphs in Feynman gauge.
Indeed, they yield leading contributions that are needed to correctly identify the standard gluon PDF. 
Calculations that go beyond lowest order in $g_s$ therefore require care in how the polarization 
of external gluon lines is treated, and involves keeping appropriate
combinations of $K$- and $G$-terms in a Grammer-Yennie 
style treatment.
Although we have obtained our result for the simple case of the integrated PDF, our arguments relied only on 
the application of Ward identities at the amplitude level.  
Therefore, similar results should hold for unintegrated PDFs 
(and related objects like fragmentation functions) 
once suitable 
definitions have been established.  Since Eq.~(\ref{eq:factKKb}) exactly vanishes in an Abelian 
theory, this result is a specific example of how the non-Abelian nature of QCD can complicate factorization 
arguments.

Higher-order calculations of hard-scattering coefficients involve subtraction terms with multiple 
target-collinear gluons, so the results obtained here are of direct importance for obtaining correct 
hard-scattering coefficients beyond lowest order 
consistent with factorization and a well-defined gluon PDF.

Standard power-counting shows that it is also possible to get leading
contributions when the hard scattering is induced by Faddeev-Popov
ghosts, but only at higher order in the hard scattering than is
treated in this paper.  It is natural to expect a generalization of
the known results \cite{Joglekar:1975nu,Henneaux:1993jn} for the
operator product expansion that applies to moments of DIS structure
functions.  On this basis we expect the general result to be that the
sum over all graphs gives extra contributions to the factorization
properties involving coefficients times what we will call alien PDFs.
The operators defining the alien PDFs are variations of certain parent
operators under Becchi-Rouet-Stora-Tyutin transformations and
therefore vanish in physical matrix elements.  Non-vanishing of the
Green functions of the alien operators with off-shell quark and gluon
states has caused calculational problems
\cite{Collins:1994ee,Harris:1994tp}.

\section*{Acknowledgments} 
We thank A. Stasto and G. Sterman for useful conversations.

Feynman diagrams 
were made using JaxoDraw~\cite{Binosi:2003yf}.
This work was supported by the
U.S. D.O.E. under grant number DE-FG02-90ER-40577.

\appendix
\section{Subtractions}
\label{sec:subtractions}

We now review the subtractive approach --- e.g., \cite{Collins:2001fm}
and Sect.\ VI of \cite{Collins:2007ph} --- we use for determining the
hard-scattering factorization.  It is generalized from the Bogoliubov
approach to renormalization.  Our strategy is to recursively examine
successively larger leading regions for the DIS cross sections.  

\begin{figure}
  \centering
  \includegraphics[scale=0.6]{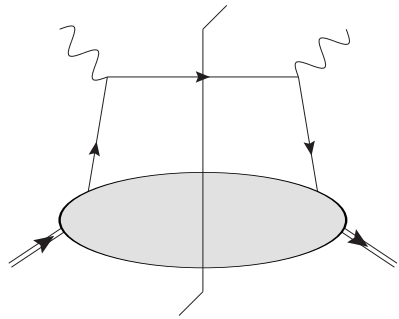}
  \caption{Handbag diagram for DIS.}
  \label{fig:handbag}
\end{figure}

\begin{figure}
  \centering
  \includegraphics[scale=0.6]{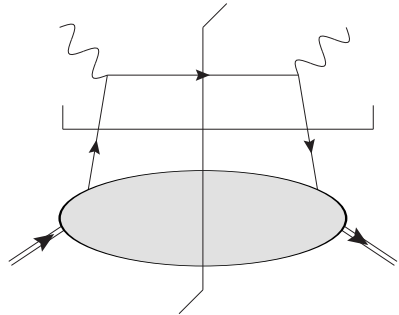}
  \caption{Parton model approximation to handbag diagram.}
  \label{fig:PM.approx}
\end{figure}

\begin{figure}
  \centering
  \includegraphics[scale=0.6]{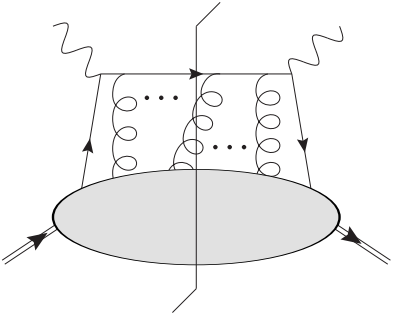}
  \caption{Generalized handbag diagram, with 
arbitrarily many extra gluon exchanges.} 
  \label{fig:extra.glue}
\end{figure}

The procedure starts with a handbag-diagram structure --- Fig.\
\ref{fig:handbag} --- to give the LO term, i.e., the
parton-model formula.  This term is obtained by an approximation valid
to leading power when the incoming quark is nearly on-shell and
collinear to the target.  The approximation is denoted by the hooked
line in Fig.\ \ref{fig:PM.approx}, and it involves replacing the
momentum of the struck quark in the upper part of the graph by its
parton-model approximation, which is massless, on-shell, and of zero
transverse momentum.

If a general graph contributing to the cross section is denoted by
$\Gamma$, then we call graphs where we make the parton-model approximation
$T_{\rm LO} \Gamma$.  At this level we also need to consider graphs with
arbitrarily many target-collinear gluons attaching to the the hard
vertex --- Fig.\ \ref{fig:extra.glue}.  A Ward identity allows the
target-collinear gluons to be disentangled from the LO hard-scattering
coefficient.  Since there are no super-leading contributions in these
graphs, and since the relevant graphs are tree graphs, the Ward
identities are unproblematic.

The result is a convolution product of the LO hard-scattering
coefficient with a sum of graphs identifiable as an expansion of the
quark PDF.  Schematically, the cross section is then written,
\begin{multline}
\label{eq:factbasicLO}
\sigma = T_{\rm LO} \sum \Gamma + \mathcal{O}(g_s^2 \sigma)
     + \mathcal{O} \!\left(\left( \frac{\Lambda}{Q} \right)^a \sigma \right) = 
\\ = \mathcal{C}_{\rm LO} \otimes f_{q/p}(Q^2) 
+ \mathcal{O}(g_s^2 \sigma) 
+ \mathcal{O} \!\left( \left( \frac{\Lambda}{Q} \right)^a \sigma \right),
\end{multline}
where $f_{q/p}(x,Q^2)$ is the quark PDF, $\mathcal{C}_{\rm LO}$ is the
LO hard-scattering coefficient (from the electromagnetic vertex), and
$\otimes$ symbolizes the usual convolution product.

Some of the errors in this approximation are power suppressed,
indicated by the term $(\Lambda/Q)^a$ where $a>0$, and we do not consider
them further.  Other errors are caused by regions with larger
transverse momentum for the quarks and gluons, and by graphs not of
the form of Figs.\ \ref{fig:handbag} and \ref{fig:extra.glue}.  These
will be covered by our treatment of higher-order scattering, and are
suppressed only by a power of the strong coupling at scale $Q$, as
indicated by the term $\mathcal{O}(g_s^2 \sigma)$.

Hence, we have the LO approximation to the DIS cross section
\begin{equation}
\label{eq:factbasicLO2}
\sigma \approx \sigma_{\rm LO} = \mathcal{C}_{\rm LO} \otimes f_{q/p}(x,Q^2).
\end{equation}

\begin{figure}
  \centering
  \begin{tabular}{c@{\hspace*{10mm}}c}
    \includegraphics[scale=0.35]{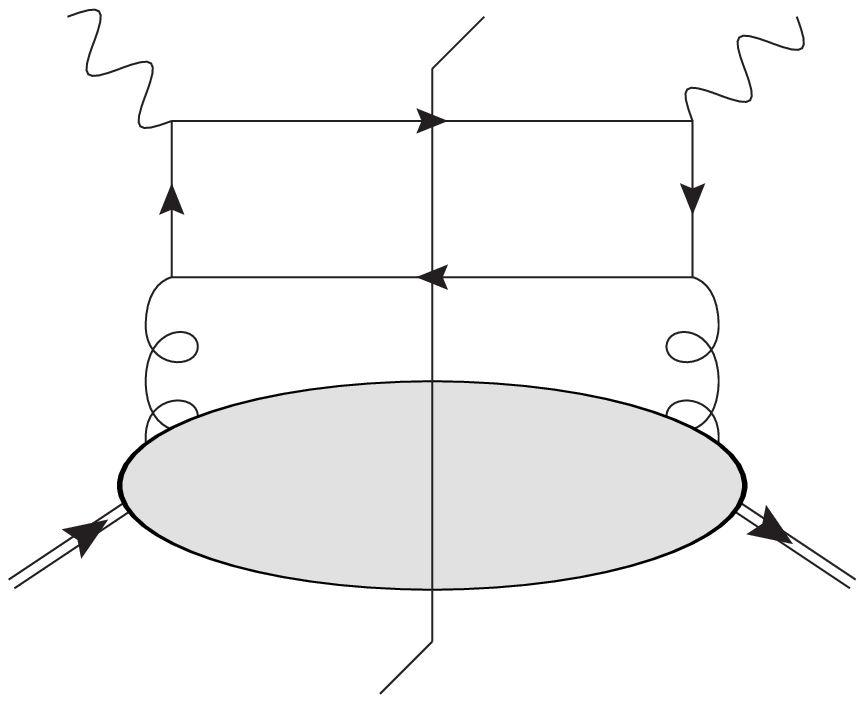}
    &
    \includegraphics[scale=0.35]{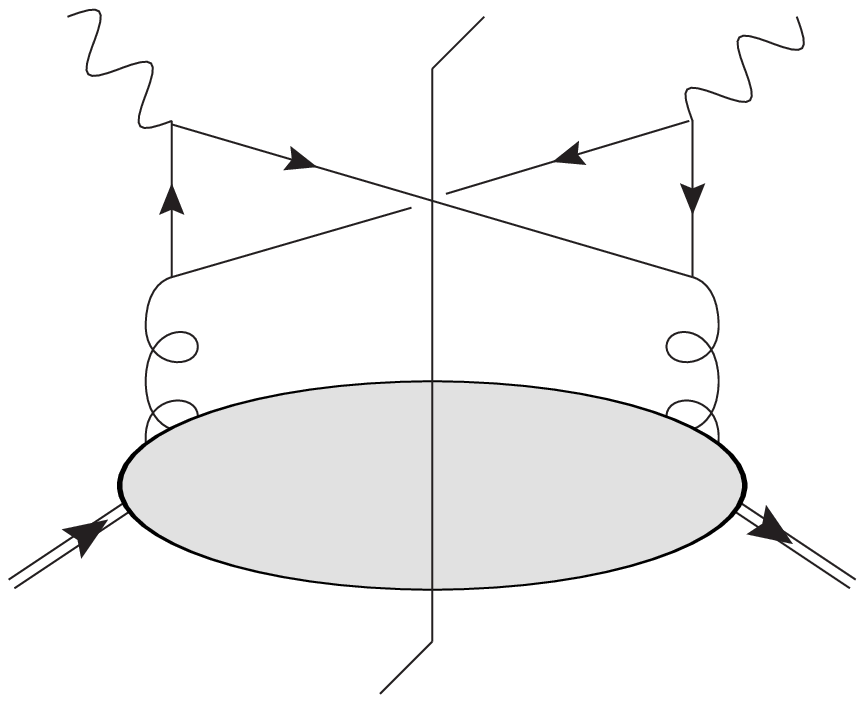}
    \\
    (a) & (b)
  \end{tabular}  
  \caption{
Examples of graphs giving NLO contributions for DIS.  
These are the relevant ones for the gluon-induced term.  But note that
there are also other graphs that give the quark-induced NLO contributions.}
  \label{fig:NLO.graphs}
\end{figure}

To find the next-to-leading (NLO) contribution, we examine, among
others, graphs of the topologies shown in Fig.\ \ref{fig:NLO.graphs}.
Specific examples are included in the graphs we treat in the main body
of the paper.  For the moment we ignore the possibility of extra gluon
exchanges.  Some of the graphs, e.g., Fig.\ \ref{fig:NLO.graphs}(a),
are among those already considered in the LO approximation, merely
viewed with a different partition between the lower bubble and the
parton lines explicitly drawn at the top.

\begin{figure}
  \centering
  \includegraphics[scale=0.35]{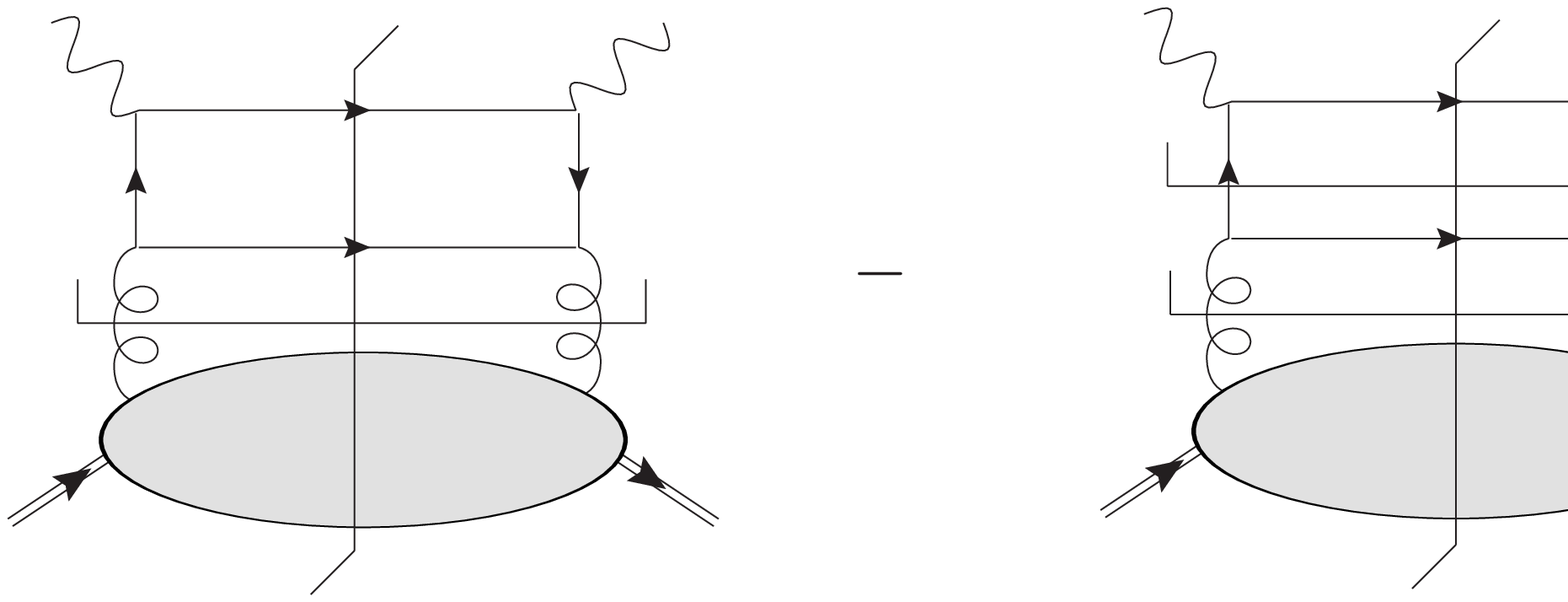}
  \caption{
Generalized parton-model approximation to Fig.\
    \ref{fig:NLO.graphs}(a).  Each graph corresponds to a term in Eq.~(\ref{eq:factbasicNLO2}).}
  \label{fig:NLO}
\end{figure}

Therefore, before we can get the true NLO contribution, we must first
subtract the LO approximation:
\begin{equation}
\label{eq:factbasicNLO}
\sum \Gamma - T_{\rm LO} \sum \Gamma.
\end{equation}
This is just the remainder from applying the approximation already
considered in Eq.~(\ref{eq:factbasicLO}).  Finally, to evaluate
Eq.~(\ref{eq:factbasicNLO}), we apply approximations that are good for
the wide-angle $2\to2$ parton subprocesses, denoted by the symbol
$T_{\rm NLO}$:
\begin{equation}
\label{eq:factbasicNLO2}
T_{\rm NLO} \!\left( \sum \Gamma - T_{\rm LO} \sum \Gamma \right) 
= \sum_j \mathcal{C}_{{\rm NLO},j} \otimes f_{j/p}(Q^2),
\end{equation} 
where we must now allow for a sum over parton flavors. For the graph
in Fig.\ \ref{fig:NLO.graphs}(a), the result is shown graphically in
Fig.\ \ref{fig:NLO}.

\begin{figure}
  \centering
   \includegraphics[scale=0.35]{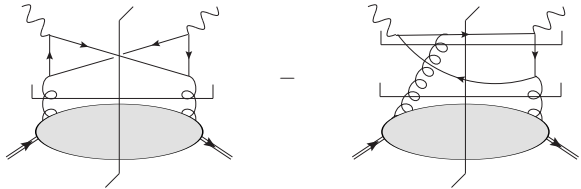}
   \caption{NLO term corresponding to Fig.\ \ref{fig:NLO.graphs}(b),
     with subtraction for non-handbag LO contribution.  The
     subtraction term uses the same graph as the first term, but it
     has been drawn differently to show that it gives a case of Fig.\
     \ref{fig:extra.glue}.}
  \label{fig:NLO1}
\end{figure}

\begin{figure}
  \centering
   \includegraphics[scale=0.4]{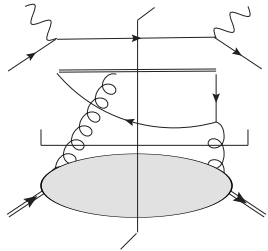}
  \caption{Particular graph for gluon-induced NLO term, with
    subtraction for non-handbag LO contribution.}
  \label{fig:NLOsub}
\end{figure}

If graphs like Fig.\ \ref{fig:NLO.graphs}(a) were all that mattered, then this 
would complete the discussion of the subtraction formalism.
However, the gauge-invariant LO term needs a generalized
handbag structure including extra gluon exchanges, as in Fig.\
\ref{fig:extra.glue}. There are thus corresponding subtractions for graphs like  Fig.\ \ref{fig:NLO.graphs}(b), as shown in
Fig.\ \ref{fig:NLO1}.  Implementing the appropriate approximations and applying 
Ward identities allows us to identify how the second term in Fig.\ \ref{fig:NLO1} 
contributes at LO.  After applying Ward identities, the second term in Fig.\ \ref{fig:NLO1} 
can be written diagrammatically as in Fig.\ \ref{fig:NLOsub}.  There we see the separation
of the graph into the LO hard-scattering matrix element and a single-gluon contribution to the 
PDF.

Thus we see that evaluating
Eq.~(\ref{eq:factbasicNLO2}) for the NLO correction requires that we
know exactly the single-gluon correction to the quark PDF in
Eq.~(\ref{eq:factbasicLO2}), since the definition of the approximation
$T_{\rm LO} \sum \Gamma$ is what gives us the lowest order, parton-model
factorization in Eq.~(\ref{eq:factbasicLO}).  In other words, using
the subtraction approach to calculate higher-order corrections to the
hard-scattering coefficient requires that we know exactly what we are
subtracting.


\end{document}